\documentclass[11pt]{article}
\usepackage{jheppub}
\pdfoutput=1
\usepackage{amsmath,bbm,array,amsfonts,graphicx,wrapfig,lscape,float,mathtools,multirow,longtable,amsthm}
\usepackage[dvipsnames]{xcolor}
\usepackage{stackrel}
\usepackage[all]{xy}
\usepackage{caption}
\usepackage{subcaption}
\usepackage[bottom]{footmisc}
\usepackage{mathrsfs}
\usepackage{tikz}
\usepackage{bm}
\usepackage{color}
\usepackage{enumerate}
\usetikzlibrary{decorations.pathreplacing}
\usepackage{diagbox}
\numberwithin{equation}{section}
\numberwithin{figure}{section}
\numberwithin{table}{section}
\usepackage{pgfplots}
\pgfplotsset{compat=1.14}
\usepackage{makecell}
\usepackage{lscape}
\usepackage[utf8]{inputenc}
\usepackage{mathtools}
\usepackage{stmaryrd}
\allowdisplaybreaks

\usepackage[autostyle=true]{csquotes}

\usepackage[toc,page]{appendix}

\usetikzlibrary{external}
\tikzset{external/system call={pdflatex \tikzexternalcheckshellescape 
		-halt-on-error
		-interaction=batchmode 
		-jobname "\image" "\texsource"
		&& pdftops -eps "\image.pdf"}}
\usepackage{tocloft}

	\title{A Survey of Toric Quivers and BPS Algebras}
	
	\author[a,b]{Jiakang Bao}
	
	\affiliation[a]{
		Department of Mathematics, City, University of London, EC1V 0HB, UK}
	\affiliation[b]{
		London Institute for Mathematical Sciences, Royal Institution, London W1S 4BS, UK}
	
	\emailAdd{jiakang.bao@city.ac.uk}
	
	\preprint{
		\begin{flushright}
			
		\end{flushright}
	}

	\abstract{In this note, we discuss some properties of the quiver BPS algebras. We consider how they would transform under different operations on the toric quivers, such as dualities and higgsing. We also give free field realizations of the algebras, in particular for the chiral quivers.
	}

\begin{document}
	\maketitle

\section{Introduction and Summary}\label{intro}
Given a toric Calabi-Yau (CY) threefold, D-branes wrapping its holomorphic cycles give rise to BPS bound states. The 4d $\mathcal{N}=1$ gauge theory can be beautifully summarized in the language of toric diagrams, quivers and brane tilings. As a realization of the BPS algebras in such supersymmetric gauge theories, quiver Yangians were introduced in \cite{Li:2020rij}. Physically, they can be derived via localization techniques in supersymmetric quantum mechanics \cite{Galakhov:2020vyb,Galakhov:2021xum}. See \cite{Yamazaki:2022cdg} for a recent summary.

Later in \cite{Galakhov:2021vbo} (see also \cite{Noshita:2021ldl,Noshita:2021dgj}), this was extended to the trigonometric and elliptic counterparts of the rational quiver Yangians. Such algebras, dubbed rational (toroidal, resp. elliptic) quiver BPS algebras, can be realized by 1d $\mathcal{N}=4$ (2d $\mathcal{N}=(2,2)$, resp. 3d $\mathcal{N}=2$) quantum field theories. These theories are low-energy effective theories on the D-branes that probe the CY threefolds. In particular, the three types of algebras can be uniformly described by some bond factors. For the elliptic algebras, the bond factor is composed of certain theta function $\Theta_q(u)$, where $q$ is the square of the nome. In other words, it is related to the modulus $\tau$ of the torus by $q=\exp(2\pi i\tau)$. Under dimensional reduction, this gives the trigonometric version of the algebras whose bond factor is determined by $\text{Sin}_{\beta}(u):=2\sinh(\beta u/2)$. In the limit where the radius $\beta$ of the circle goes to 0, one reaches the rational case with the bond factor consisting of the rational function $u$. All these algebras have two parameters $h_{1,2}$.

On the other hand, it is well-known that for supersymmetric gauge theories on toric CYs, many features, such as dualities and higgsing, can be nicely described using quivers and brane tilings \cite{Feng:2000mi,Feng:2001xr,Beasley:2001zp,Feng:2001bn,Feng:2002zw,Feng:2002fv,Franco:2005rj}. It is then natural to ask what properties the quiver BPS algebras would have under these features.

As each quiver has its associated quiver BPS algebras, it is conjectured that the corresponding quiver BPS algebras are isomorphic under Seiberg/toric duality\footnote{In general, Seiberg duality can also take quivers outside the toric phases, leading to the phenomenon called duality cascades \cite{Klebanov:2000hb,Franco:2003ja}. One can still define the algebras for these quivers following \S\ref{algebras}. However, whether they would give rise to the corresponding BPS algebras needs further study. Therefore, we shall only focus on toric quivers here.}. In the case of rational quiver Yangians for toric CY threefolds without compact divisors, this was proven in \cite{Bao:2022jhy}. Here, we shall have a discussion on the other cases.

For toroidal algebras from non-chiral quivers, we can construct the transformations of the generators under toric duality that are similar to the rational cases. As such construction is based on the modes of the algebra, this approach becomes more difficult in the elliptic cases as the current relations involve $q$-Pochhammer symbols. Therefore, we would like to work with the current relations. There are four types of currents, $e$, $f$ and $\psi_{\pm}$. As shown in \cite{Galakhov:2021vbo}, all the other current relations can be derived from the $ee$, $ff$ and $ef$ relations. Hence, it is sufficient to consider these relations to construct the transformations for dual algebras. Suppose the node $\digamma$ is dualized. Although the elliptic cases are rather involved, we propose that the transformations of the currents associated to $\digamma\pm1$ can be determined by the corresponding ``brackets'' of the three types of algebras.

For chiral quivers that are associated to toric CYs with compact divisors, there do not seem to have underlying Kac-Moody superalgebras for the quiver BPS algebras. The patterns of the mode relations would also vary from case to case. As a result, it is not easy to study the connections of toric dual algebras even for the rational quiver Yangian case. As they are quite different from the non-chiral cases even if we directly consider the current relations, we shall only give a naive construction for the currents here. As a result, this would only give a valid transformation in the trivial case where the two parameters of the algebra vanish. A proof of isomorphisms for general cases might require more sophisticated methods, and it might also be helpful to first start with some specific examples. Nevertheless, we hope that the discussions here would provide some basic ideas of constructing such transformations.

Although the exact maps are still not known in general, we can summarize some common features for both chiral and non-chiral quivers. Suppose that the node $\digamma$ is dualized. Roughly speaking, the roles of $e^{(\digamma)}$ and $f^{(\digamma)}$ should get swapped under toric duality. For $\psi^{(\digamma)}_{\pm}(u)$, we expect them to become their inverses $\psi^{(\digamma)}_{\pm}\left(\widetilde{u}\right)^{-1}$. This is not surprising as the arrows connected to $\digamma$ would get reversed under toric duality. Moreover, for nodes that are connected to the dualized node, their associated generators should always be combined with certain generators for $\digamma$ under toric duality.

The transformations of the toric dual algebras would also shed light on the discussions related to the Higgs-Kibble mechanism. Mathematically, this corresponds to blowing up/down the singularities in the toric setting. In the rational and toroidal cases for non-chiral quivers, when they degenerate to one-parameter algebras, there is a surjection of the algebras from the parent theory to the higgsed one. The construction is similar to the one discussed in toric duality. We conjecture that higgsing would still give the subalgebra structure for general cases (at least in the one-parameter degeneracy).

The paper is organized as follows. In \S\ref{algebras}, after defining the quiver BPS algebras, we give some properties of the algebras, including coproducts and gradings. In \S\ref{toroidal}, we consider how the toroidal algebras would transform under toric duality and higgsing for non-chiral quivers. We then have some discussions on the elliptic cases, as well as the algebras for chiral quivers, in \S\ref{ellipticandchiral}. In \S\ref{discussions}, we mention some prospects regarding specular duality, truncations and vertex operator algebras. We review some basic aspects of quiver gauge theories in Appendix \ref{gauge}. In Appendix \ref{serre}, we list the Serre relations for algebras associated to (most) non-chiral quivers. In Appendix \ref{kmodes}, we make some supplementary comments on the modes of the quiver BPS algebras.

\section{Quiver BPS Algebras}\label{algebras}
The quiver BPS algebras are generated by three types of currents, $\psi^{(a)}_{\pm}(u)$, $e^{(a)}(u)$ and $f^{(a)}(u)$, where $a$ denotes the nodes of a given quiver. When there is an odd number of adjoint loops on a node $a$, we say that it is bosonic with $|a|=0$. Otherwise, it is fermionic with $|a|=1$. This naturally endows the algebras with a $\mathbb{Z}_2$-grading such that $\left|e^{(a)}(u)\right|=\left|f^{(a)}(u)\right|=|a|$ while $\psi^{(a)}_{\pm}(u)$ are always bosonic. The currents have the following mode expansions\footnote{One can also consider shifted quiver BPS algebras that would introduce an extra shift parameter to (some part of) the mode expansion of $\psi_{\pm}$ \cite{Galakhov:2021vbo}. This is closely related to the crystal representations and the framings of the quivers \cite{Galakhov:2021xum,Noshita:2021dgj}. However, we shall not consider this here.}:
\begin{equation}
	x^{(a)}(u)=\begin{cases}
		\sum\limits_{n\in\mathbb{Z}_+}\frac{x^{(a)}_n}{u^n},&\text{rational}\\
		\sum\limits_{n\in\mathbb{Z}}\frac{x^{(a)}_n}{U^n},&\text{trigonometric}\\
		\sum\limits_{n\in\mathbb{Z}}\frac{x^{(a)}_n}{U^n}=\sum\limits_{n\in\mathbb{Z}}\sum\limits_{\alpha\in\mathbb{Z}_{\geq0}}\frac{x^{(a)}_{n,\alpha}}{U^n}q^\alpha,&\text{elliptic}
	\end{cases},
\end{equation}
where $x=e,f$. For non-chiral quivers,
\begin{equation}
	\psi^{(a)}_{\pm}(u)=\begin{cases}
		\sum\limits_{n\in\mathbb{Z}_{\geq0}}\frac{\psi^{(a)}_n}{u^n},&\text{rational}\\
		\sum\limits_{n\in\mathbb{Z}_{\geq0}}\frac{\psi^{(a)}_{\pm,n}}{U^{\pm n}},&\text{trigonometric}\\
		\sum\limits_{n\in\mathbb{Z}}\frac{\psi^{(a)}_{\pm,n}}{U^{\pm n}}=\sum\limits_{n\in\mathbb{Z}}\sum\limits_{\alpha\in\mathbb{Z}_{\geq0}}\frac{\psi^{(a)}_{\pm,n,\alpha}}{U^{\pm n}}q^\alpha,&\text{elliptic}
	\end{cases}
\end{equation}
with $\psi^{(a)}_{\pm,n<0,0}=0$ in the elliptic case. For chiral quivers,
\begin{equation}
	\psi^{(a)}_{\pm}(u)=\begin{cases}
		\sum\limits_{n\in\mathbb{Z}}\frac{\psi^{(a)}_n}{u^n},&\text{rational}\\
		\sum\limits_{n\in\mathbb{Z}}\frac{\psi^{(a)}_{\pm,n}}{U^{\pm n}},&\text{trigonometric}\\
		\sum\limits_{n\in\mathbb{Z}}\frac{\psi^{(a)}_{\pm,n}}{U^{\pm n}}=\sum\limits_{n\in\mathbb{Z}}\sum\limits_{\alpha\in\mathbb{Z}_{\geq0}}\frac{\psi^{(a)}_{\pm,n,\alpha}}{U^{\pm n}}q^\alpha,&\text{elliptic}
	\end{cases}.
\end{equation}
Notice that in the rational case, $\psi_+=\psi_-=\psi$. Moreover, the expansions for trigonometric and elliptic cases are in terms of $U$ rather than $u$. The letters in the upper case are related to those in the lower case by\footnote{As a result, $\beta$ can be absorbed under a redefinition of variables. Nevertheless, we shall keep it here due to its physical origin.}
\begin{equation}
	X=\text{e}^{\beta x},\qquad (x,X)=(u,U),(v,V),(c,C),(\mathtt{h}_I,\mathtt{H}_I),\dots
\end{equation}
Henceforth, we will use the upper and lower cases interchangeably (such as $e^{(a)}(U)=e^{(a)}(u)$) in the arguments of the currents for trionometric and elliptic cases. For convenience, we may also write $e^{(a)}(U)=\sum\limits_{\alpha\in\mathbb{Z}_{\geq0}}e^{(a)}_\alpha(U)q^\alpha$ (and likewise for the other currents) in the elliptic case.

To write their relations, we first need to introduce some necessary concepts. To distinguish chiral and non-chiral quivers, we define the chirality parameter as
\begin{equation}
	\chi_{ab}=|a\rightarrow b|-|b\rightarrow a|
\end{equation}
for each pair of nodes $a$, $b$ in the quiver, where $|a\rightarrow b|$ denotes the number of arrows from $a$ to $b$. Moreover, we shall define the formal delta function as
\begin{equation}
	\delta(u)=\begin{cases}
		1/u,&\text{rational}\\
		\sum\limits_{n\in\mathbb{Z}}U^n,&\text{otherwise}
	\end{cases}.
\end{equation}

The key factor in the definition of the algebras is the bond factor $\varphi^{a\Leftarrow b}(u)$. In this paper, we shall write it as\footnote{This is slightly different from the notion in \cite{Galakhov:2021vbo} when both $|a\rightarrow b|$ and $|b\rightarrow a|$ are odd. Nevertheless, the bond factor here should still be legitimate as it satisfies the reciprocity condition \eqref{reciprocity} below.}
\begin{equation}
	\varphi^{a\Leftarrow b}(u)=\frac{\prod\limits_{a\rightarrow b}\zeta(\mathtt{h}_I+u)}{\prod\limits_{b\rightarrow a}\zeta(\mathtt{h}_I-u)},
\end{equation}
where $\mathtt{h}_I$ is the parameter/charge associated to the arrow $I$ in the quiver, and
\begin{equation}
	\zeta(u)=\begin{cases}
		u,&\text{rational}\\
		\text{Sin}_\beta(u):=2\sinh\frac{\beta u}{2}=U^{1/2}-U^{-1/2},&\text{trigonometric}\\
		\Theta_q(u):=-U^{-1/2}\theta_q(u)=\left(U^{1/2}-U^{-1/2}\right)\prod\limits_{k=1}^{\infty}\left(1-U^{-1}q^k\right)(1-Uq^k),&\text{elliptic}
	\end{cases}.
\end{equation}
Here, $\theta_q(u)=\left(U;q\right)_{\infty}\left(qU^{-1};q\right)_{\infty}$ in terms of the $q$-Pochhammer symbols. From the expressions for $\zeta$, we have
\begin{equation}
	\zeta(u)=-\zeta(-u).
\end{equation}
It is straightforward to see that in the rational limit $\beta\rightarrow0$, the trigonometric case reduces to the rational one. Likewise, when $q\rightarrow0$, the elliptic one reduces to the trigonometric one. This will also be the limits that relate the three types of quiver BPS algebras.

To get rid of the powers with half-integers, we will take the balanced bond factor
\begin{equation}
	\phi^{a\Leftarrow b}(u,v)=(UV)^{\frac{\mathfrak{t}}{2}\chi_{ab}}\varphi^{a\Leftarrow b}(u-v),\label{balancing}
\end{equation}
where $\mathfrak{t}$ is 0 in the rational case and $-1$ otherwise\footnote{Notice that this is slightly different from the original one in \cite{Galakhov:2021vbo} where $\mathfrak{t}$ was defined to be 1 for the trigonometric and elliptic cases. This is only a choice for our later discussions on mode expansions for chiral quivers. Since the balancing factor $(UV)^{\frac{\mathfrak{t}}{2}\chi_{ab}}$ is used to get rid of the half-integer powers in the Laurent expansions of the expressions, this should just be a matter of convention.}. Therefore, this balancing would only affect chiral quivers in the trigonometric and elliptic cases. As can be seen from its expression, the bond factor satisfies
\begin{equation}
	\varphi^{a\Leftarrow b}(u)\varphi^{b\Leftarrow a}(-u)=1.\label{reciprocity}
\end{equation}
Therefore,
\begin{equation}
	\phi^{a\Leftarrow b}(u,v)\phi^{b\Leftarrow a}(v,u)=1.
\end{equation}
Moreover, we have
\begin{equation}
	\phi^{a\Leftarrow b}(u+s,v)=s^{\mathfrak{t}\chi_{ab}}\phi^{a\Leftarrow b}(u,v-s).
\end{equation}

With this (balanced) bond factor, the three types of quiver Yangians can be presented in a unified way as \cite{Li:2020rij,Galakhov:2021vbo}
\begin{align}
	&\psi^{(a)}_{\pm}(u)\psi^{(b)}_{\pm}(v)\simeq C^{\pm\mathfrak{t}\chi_{ab}}\psi^{(b)}_{\pm}(v)\psi^{(a)}_{\pm}(u),\label{psipsi1}\\
	&\psi^{(a)}_+(u)\psi^{(b)}_-(v)\simeq\frac{\phi^{a\Leftarrow b}(u+c/2,v-c/2)}{\phi^{a\Leftarrow b}(u-c/2,v+c/2)}\psi^{(b)}_-(v)\psi^{(a)}_+(u),\\
	&\psi^{(a)}_{\pm}(u)e^{(b)}(v)\simeq\phi^{a\Leftarrow b}(u\pm c/2,v)e^{(b)}(v)\psi^{(a)}_{\pm}(u),\\
	&\psi^{(a)}_{\pm}(u)f^{(b)}(v)\simeq\phi^{a\Leftarrow b}(u\mp c/2,v)^{-1}f^{(b)}(v)\psi^{(a)}_{\pm}(u),\label{psif}\\
	&e^{(a)}(u)e^{(b)}(v)\simeq(-1)^{|a||b|}\phi^{a\Leftarrow b}(u,v)e^{(b)}(v)e^{(a)}(u),\label{ee}\\
	&f^{(a)}(u)f^{(b)}(v)\simeq(-1)^{|a||b|}\phi^{a\Leftarrow b}(u,v)^{-1}f^{(b)}(v)f^{(a)}(u),\\
	&\left[e^{(a)}(u),f^{(b)}(v)\right\}\simeq-\delta_{ab}\left(\delta(u-v-c)\psi^{(a)}_+(u-c/2)-\delta(u-v+c)\psi^{(a)}_-(v-c/2)\right)\label{ef}.
\end{align}
Here, $c$ is the central element of the algebra which is 0 for the rational case (while it can be non-trivial for the other two cases). In the last relation, we have used the supercommutator $[x,y\}=xy-(-1)^{|x||y|}yx$. For the rational quiver Yangians, ``$\simeq$'' indicates that the equalities are up to some sporadic $u^mv^n$ terms\footnote{This is to ensure that the mode relations, which can be found for example in (4.20) in \cite{Li:2020rij}, would be exact.}. For the trigonometric and elliptic cases, it means that the Laurent expansion on the two sides should agree, and we shall henceforth simply write it as ``$=$''. As shown in \cite{Galakhov:2021vbo}, \eqref{psipsi1}$\sim$\eqref{psif} can be derived from \eqref{ee}$\sim$\eqref{ef}. Therefore, when discussing the current relations, it suffices to consider the $ee$, $ff$ and $ef$ relations.

Besides the above relations, we also need the Serre relations. It is believed that the Serre relations are closely related to the superpotential although the precise relations are still not known in general\footnote{Note added in version 2: Very recently, the Serre relations for any quivers in the toroidal case were found in \cite{Negut:2023iia}. In the rational limit, they should give rise to the Serre relations in the rational case, and they should also yield those for the elliptic algebras using the dressed operators in Appendix \ref{serre}. We expect that our following discussions would be compatible with these Serre relations.}. For generalized conifolds and some chiral quivers, their Serre relations were given in \cite{Li:2020rij,Galakhov:2021vbo}. See also Appendix \ref{serre}.

In general, these quiver BPS algebras are two-parameter algebras. This is because the charges $\mathtt{h}_I$ should satisfy the loop and vertex constraints. Due to the uniquely determined superpotential in the toric setting, each of its monomial term $L$, whose arrows form a closed loop in the quiver, yields one loop constraint: $\sum\limits_{I\in L}\mathtt{h}_I=0$. Moreover, the algebra also has shift automorphisms $\mathtt{h}_I\rightarrow \mathtt{h}_I+\text{sgn}_a(I)\epsilon_a$ for some parameters $\epsilon_a$, where $\text{sgn}_a(I)$ is 1 ($-1$, resp. 0) if $I\in\{a\rightarrow b|b\neq a\}$ ($I\in\{b\rightarrow a|b\neq a\}$, resp. otherwise). We can use the vertex constraint $\sum\limits_I\text{sgn}_a\mathtt{h}_I=0$ to mod out these gauge symmetry redundancies. Overall, we have two free parameters, say $h_1$ and $h_2$, left. Together with the R-symmetry, they give the U$(1)^3$ isometry of the toric CY threefold.

\subsection{Coproducts}\label{coprod}
The coproducts of the quiver BPS algebras are of particular interest due to their crucial role in the construction of $R$-matrices and the study of Bethe/gauge correspondence \cite{Galakhov:2022uyu,Bao:2022fpk}. For rational quiver Yangians of certain non-chiral quivers, the coproduct was given in \cite{Bao:2022jhy} using the techniques developed in \cite{guay2018coproduct}. However, this is still not known for chiral quivers\footnote{The reason is that the quiver Yangians for chiral quivers do not seem to have underlying Kac-Moody superalgebras which are quite heavily relied on when finding the coproduct for the non-chiral quiver cases. Due to the complication in the current relations, we also need to write the coproduct in terms of modes, which is more difficult.}. In contrast, the coproducts for trigonometric and elliptic cases are more straightforward. One may verify that the following gives a coassociative homomorphism (cf. \cite{Noshita:2021ldl}):
\begin{align}
	&\Delta\left(e^{(a)}(U)\right)=e^{(a)}(U)\otimes1+\psi^{(a)}\left(C_1^{1/2}U\right)\otimes e^{(a)}(C_1U),\\
	&\Delta\left(f^{(a)}(U)\right)=1\otimes f^{(a)}(U)+f^{(a)}(C_2U)\otimes f^{(a)}\left(C_2^{1/2}U\right),\\
	&\Delta\left(\psi^{(a)}_+(U)\right)=\psi^{(a)}_+(U)\otimes\psi^{(a)}_+\left(C_1^{-1}U\right),\\
	&\Delta\left(\psi^{(a)}_-(U)\right)=\psi^{(a)}_-\left(C_2^{-1}U\right)\otimes\psi^{(a)}_-(U),\\
	&\Delta(C)=C\otimes C.
\end{align}
Here, $C_1=C\otimes1$ and $C_2=1\otimes C$ are the conventional notations that indicate where the $C$ factors should be in the mode expressions. More explicitly, for the toroidal algebras associated to non-chiral quivers, we have
\begin{align}
	&\Delta\left(e^{(a)}_n\right)=e^{(a)}_n\otimes1+\sum_{j=0}^{\infty}C^{-n-j/2}\psi^{(a)}_{-,j}\otimes e^{(a)}_{n+j},\\
	&\Delta\left(f^{(a)}_n\right)=1\otimes f^{(a)}_n+\sum_{j=0}^{\infty}f^{(a)}_{n-j}\otimes C^{-n+j/2}\psi^{(a)}_{+,j},\\
	&\Delta\left(\psi^{(a)}_{+,n}\right)=\sum_{j=0}^nC^{n-j}\psi^{(a)}_{+,j}\otimes\psi^{(a)}_{+,n-j},\\
	&\Delta\left(\psi^{(a)}_{-,n}\right)=\sum_{j=0}^n\psi^{(a)}_{-,n-j}\otimes C^{-n+j}\psi^{(a)}_{-,j}.
\end{align}
For the elliptic algebras and the algebras for chiral quivers, we just need to replace all $\sum\limits_{j=0}^{\infty}$ and $\sum\limits_{j=0}^n$ with $\sum\limits_{j\in\mathbb{Z}}$. In the elliptic case, it is also straightforward to write down this in terms of $x^{(a)}_{n,\alpha}$ ($x=e,f,\psi_{\pm}$). We simply replace $x_n\otimes 1$ (resp. $1\otimes x_n$) with $x_{n,\alpha}\otimes1$ (resp. $1\otimes x_{n,\alpha}$) and $x_m\otimes y_n$ with $\sum\limits_{\gamma=0}^{\alpha}x_{m,\gamma}\otimes y_{n,\alpha-\gamma}$.

\paragraph{Hopf algebras} Together with the above coproduct in terms of the currents, we can have a counit and an antipode such that the algebra is endowed with the Hopf (super)algebra structure. The counit reads
\begin{equation}
	\epsilon\left(e^{(a)}(U)\right)=\epsilon\left(f^{(a)}(U)\right)=0,\quad\epsilon\left(\psi^{(a)}_{\pm}(U)\right)=\epsilon(C)=1.
\end{equation}
The antipode is an anti-homomorphism, that is, $S(xy)=(-1)^{|x||y|}S(y)S(x)$. Assuming that $\psi^{(a)}_{\pm}(U)$ are invertible in the algebra, then
\begin{align}
	&S\left(e^{(a)}(U)\right)=-\psi^{(a)}_-\left(C^{-3/2}U\right)^{-1}e^{(a)}(CU),\\
	&S\left(f^{(a)}(U)\right)=-f^{(a)}(CU)\psi^{(a)}_+\left(C^{-3/2}U\right)^{-1},\\
	&S\left(\psi^{(a)}_{\pm}(U)\right)=\psi^{(a)}_{\pm}\left(C^{-1}U\right)^{-1},\\
	&S(C)=C^{-1}.
\end{align}
It is also straightforward to write them in terms of the modes. One may check that they satisfy the properties of Hopf algebras, such as $\mathcal{M}\circ(S\times\text{id})\circ\Delta=\mathcal{M}\circ(\text{id}\times S)\circ\Delta=\eta\circ\epsilon$, where $\mathcal{M}$ and $\eta$ denote the multiplication and the unit map respectively.

\subsection{Gradings}\label{gradings}
Similar to the discussions in \cite{feigin2015quantum,garbali2021r}, we can assign different gradings to the quiver BPS algebras. The degree of an element $x$ can be written as $\text{deg}(x)=(\text{pdeg}(x),\text{hdeg}(x))$, where $\text{pdeg}(x)=\left(\text{pdeg}^{(a)}(x)\right)$ is a vector known as the principal degree and $\text{hdeg}(x)$ is a number called the homogeneous degree. We can introduce some invertible elements $D^{(a)}$ and $D$ such that $D^{(a)}x\left(D^{(a)}\right)^{-1}=\text{e}^{\beta\text{pdeg}^{(a)}(x)}x$ and $DxD^{-1}=\text{e}^{-\beta\text{hdeg}(x)}x$.

We have $\text{deg}\left(e^{(a)}_n\right)=(0,\dots,1,\dots,0,n)$ and $\text{deg}\left(f^{(a)}_n\right)=(0,\dots,-1,\dots,0,n)$, where $\pm1$ is at the $a^{\text{th}}$ entry. On the other hand, $\text{deg}\left(\psi^{(a)}_{\pm,n}\right)=(\bm{0},\pm n)$ and $\text{deg}(C)=\text{deg}\left(D^{(a)}\right)=\text{deg}(D)=(\bm{0},0)$. For elliptic algebras, we may further consider the degree with respect to $q$, as well as an operator $D_q$, such that the modes at order $\alpha$ would have $q$-deg equal to $\alpha$.

\section{Toroidal Algebras for Non-Chiral Quivers}\label{toroidal}
The first examples we shall discuss are the toroidal algebras for non-chiral quivers. Here, we will mainly focus on the generalized conifolds $xy=z^Mw^N$ with $M+N\geq3$. In particular, it suffices to consider these cases in the discussions of toric duality as the other cases all have one single toric phase\footnote{Of course, for $M+N\geq3$, all the triangles (i.e., $MN=0$) and the suspended pinch point (i.e., $(M,N)=(2,1),(1,2)$) have one single toric phase as well.}.

We shall use the same convention as in \cite{Galakhov:2021vbo} for the two parameters $h_{1,2}$ of the algebra. For the arrow pointing from $a$ to $b$, its charge is
\begin{equation}
	h_{ab}=A_{ab}h_1+M_{ab}h_2=\begin{cases}
		2\varsigma_ah_1,&a=b\\
		\varsigma_b(-h_1-h_2),&a+1=b\\
		\varsigma_a(-h_1+h_2),&a=b+1\\
		0,&\text{otherwise}
	\end{cases},
\end{equation}
where the definition of the signs $\varsigma_a$, as well as the construction of the quivers from a given toric diagram, can be found in Appendix \ref{gencon}. Equivalently, we can write $H_{ab}=H_1^{A_{ab}}H_2^{M_{ab}}$. Here, $A_{ab}$ is the Cartan matrix
\begin{equation}
	A_{ab}=(\varsigma_a+\varsigma_{a+1})\delta_{ab}-\varsigma_a\delta_{a,b+1}-\varsigma_b\delta_{a+1,b},
\end{equation}
and $M_{ab}$ is defined as
\begin{equation}
	M_{ab}=\varsigma_a\delta_{a,b+1}-\varsigma_b\delta_{a+1,b}.
\end{equation}
Therefore, $A_{ab}$ is symmetric while $M_{ab}$ is antisymmetric.

The relations for the toroidal quiver algebra $\mathtt{T}$ can then be explicitly written as
\begin{align}
	&\psi^{(a)}_{\pm}(U)\psi^{(b)}_{\pm}(V)=\psi^{(b)}_{\pm}(V)\psi^{(a)}_{\pm}(U),\label{toroidalglmnpsipsi1}\\
	&\frac{H_2^{M_{ab}}H_1^{A_{ab}}U-CV}{H_2^{M_{ab}}U-H_1^{A_{ab}}CV}\psi^{(a)}_{\pm}(U)\psi^{(b)}_{\mp}(V)=\frac{H_2^{M_{ab}}H_1^{A_{ab}}CU-V}{H_2^{M_{ab}}CU-H_1^{A_{ab}}V}\psi^{(b)}_{\mp}(V)\psi^{(a)}_{\pm}(U),\\
	&\left(H^{M_{ab}}C^{\pm1/2}U-H_1^{A_{ab}}V\right)\psi^{(a)}_{\pm}(U)e^{(b)}(V)=\left(H_2^{M_{ab}}H_1^{A_{ab}}C^{\pm1/2}U-V\right)e^{(b)}(V)\psi^{(a)}_{\pm}(U),\\
	&\left(H^{M_{ab}}C^{\mp1/2}U-H_1^{-A_{ab}}V\right)\psi^{(a)}_{\pm}(U)f^{(b)}(V)=\left(H_2^{M_{ab}}H_1^{-A_{ab}}C^{\mp1/2}U-V\right)f^{(b)}(V)\psi^{(a)}_{\pm}(U),\\
	&\left(H^{M_{ab}}U-H_1^{A_{ab}}V\right)e^{(a)}(U)e^{(b)}(V)=(-1)^{|a||b|}\left(H_2^{M_{ab}}H_1^{A_{ab}}U-V\right)e^{(b)}(V)e^{(a)}(U),\\
	&\left(H^{M_{ab}}U-H_1^{-A_{ab}}V\right)f^{(a)}(U)f^{(b)}(V)=(-1)^{|a||b|}\left(H_2^{M_{ab}}H_1^{-A_{ab}}U-V\right)f^{(b)}(V)f^{(a)}(U),\\
	&\left[e^{(a)}(U),f^{(b)}(V)\right\}=-\delta_{ab}\left(\delta\left(UV^{-1}C^{-1}\right)\psi^{(a)}_+\left(UC^{-1/2}\right)-\delta\left(UV^{-1}C\right)\psi^{(a)}_-\left(VC^{-1/2}\right)\right)\label{toroidalglmnef}.
\end{align}
The Serre relations are given in Appendix \ref{serre}. Moreover, when the central charge is trivial, that is, when $C=1$, $\psi_+$ would commute with $\psi_-$ as can be seen directly from their current relations.

\subsection{More on Mode Expansions}\label{modes}
We can also express \eqref{toroidalglmnpsipsi1}$\sim$\eqref{toroidalglmnef} in terms of modes:
\begin{align}
	&\psi^{(a)}_{\pm,m}\psi^{(b)}_{\pm,n}=\psi^{(b)}_{\pm,n}\psi^{(a)}_{\pm,m},\\
	&\nonumber\\
	&~~H_2^{2M_{ab}}H_1^{A_{ab}}C\psi^{(a)}_{+,m+2}\psi^{(b)}_{-,n}-H_2^{M_{ab}}\left(H_1^{2A_{ab}}+C^2\right)\psi^{(a)}_{+,m+1}\psi^{(b)}_{-,n-1}-H_1^{A_{ab}}C\psi^{(a)}_{+,m}\psi^{(b)}_{-,n-2}\nonumber\\
	&=H_2^{2M_{ab}}H_1^{A_{ab}}C\psi^{(b)}_{-,n}\psi^{(a)}_{+,m+2}-H_2^{M_{ab}}\left(H_1^{2A_{ab}}C^2+1\right)\psi^{(b)}_{-,n-1}\psi^{(a)}_{+,m+1}-H_1^{A_{ab}}C\psi^{(b)}_{-,n-2}\psi^{(a)}_{+,m},\\
	&\nonumber\\
	&H_2^{M_{ab}}C^{1/2}\psi^{(a)}_{+,m+1}e^{(b)}_n-H_1^{A_{ab}}\psi^{(a)}_{+,m}e^{(b)}_{n+1}=H_2^{M_{ab}}H_1^{A_{ab}}C^{1/2}e^{(b)}_n\psi^{(a)}_{+,m+1}-e^{(b)}_{n+1}\psi^{(a)}_{+,m},\\
	&H_2^{M_{ab}}C^{-1/2}\psi^{(a)}_{-,m}e^{(b)}_n-H_1^{A_{ab}}\psi^{(a)}_{-,m+1}e^{(b)}_{n+1}=H_2^{M_{ab}}H_1^{A_{ab}}C^{-1/2}e^{(b)}_n\psi^{(a)}_{-,m}-e^{(b)}_{n+1}\psi^{(a)}_{-,m+1},\\
	&H_2^{M_{ab}}C^{-1/2}\psi^{(a)}_{+,m+1}f^{(b)}_n-H_1^{-A_{ab}}\psi^{(a)}_{+,m}f^{(b)}_{n+1}=H_2^{M_{ab}}H_1^{-A_{ab}}C^{-1/2}f^{(b)}_n\psi^{(a)}_{+,m+1}-f^{(b)}_{n+1}\psi^{(a)}_{+,m},\\
	&H_2^{M_{ab}}C^{1/2}\psi^{(a)}_{-,m}f^{(b)}_n-H_1^{-A_{ab}}\psi^{(a)}_{-,m+1}f^{(b)}_{n+1}=H_2^{M_{ab}}H_1^{-A_{ab}}C^{1/2}f^{(b)}_n\psi^{(a)}_{-,m}-f^{(b)}_{n+1}\psi^{(a)}_{-,m+1},\\
	&H_2^{M_{ab}}e^{(a)}_{m+1}e^{(b)}_n-H_1^{A_{ab}}e^{(a)}_{m}e^{(b)}_{n+1}=(-1)^{|a||b|}\left(H_2^{M_{ab}}H_1^{A_{ab}}e^{(b)}_ne^{(a)}_{m+1}-e^{(b)}_{n+1}e^{(a)}_{m}\right),\\
	&H_2^{M_{ab}}f^{(a)}_{m+1}f^{(b)}_n-H_1^{-A_{ab}}f^{(a)}_{m}f^{(b)}_{n+1}=(-1)^{|a||b|}\left(H_2^{M_{ab}}H_1^{-A_{ab}}f^{(b)}_nf^{(a)}_{m+1}-f^{(b)}_{n+1}f^{(a)}_{m}\right),\\
	&\left[e^{(a)}_m,f^{(b)}_n\right\}=-\delta_{ab}\left(C^{(m-n)/2}\psi^{(a)}_{+,m+n}-C^{-(m-n)/2}\psi^{(a)}_{-,-m-n}\right).
\end{align}
Notice that $\psi_{\pm,l<0}$ simply vanishes such as in the $ef$ relations. In particular, the $\psi e$ and $\psi f$ relations include
\begin{equation}
	\psi^{(a)}_{\pm,0}e^{(b)}_n=H_1^{\pm A_{ab}}e^{(b)}_n\psi^{(a)}_{\pm,0},\quad\psi^{(a)}_{\pm,0}f^{(b)}_n=H_1^{\mp A_{ab}}f^{(b)}_n\psi^{(a)}_{\pm,0}\label{psi0ef}
\end{equation}
by setting $m=-1$ and
\begin{equation}
	\begin{split}
		&H_2^{M_{ab}}C^{1/2}\psi^{(a)}_{+,1}e^{(b)}_n-H_1^{A_{ab}}\psi^{(a)}_{+,0}e^{(b)}_{n+1}=H_2^{M_{ab}}H_1^{A_{ab}}C^{1/2}e^{(b)}_n\psi^{(a)}_{+,1}-e^{(b)}_{n+1}\psi^{(a)}_{+,0},\\
		&H_2^{M_{ab}}C^{-1/2}\psi^{(a)}_{-,0}e^{(b)}_n-H_1^{A_{ab}}\psi^{(a)}_{-,1}e^{(b)}_{n+1}=H_2^{M_{ab}}H_1^{A_{ab}}C^{-1/2}e^{(b)}_n\psi^{(a)}_{-,0}-e^{(b)}_{n+1}\psi^{(a)}_{-,1},\\
		&H_2^{M_{ab}}C^{-1/2}\psi^{(a)}_{+,1}f^{(b)}_n-H_1^{-A_{ab}}\psi^{(a)}_{+,0}f^{(b)}_{n+1}=H_2^{M_{ab}}H_1^{-A_{ab}}C^{-1/2}f^{(b)}_n\psi^{(a)}_{+,1}-f^{(b)}_{n+1}\psi^{(a)}_{+,0},\\
		&H_2^{M_{ab}}C^{1/2}\psi^{(a)}_{-,0}f^{(b)}_n-H_1^{-A_{ab}}\psi^{(a)}_{-,1}f^{(b)}_{n+1}=H_2^{M_{ab}}H_1^{-A_{ab}}C^{1/2}f^{(b)}_n\psi^{(a)}_{-,0}-f^{(b)}_{n+1}\psi^{(a)}_{-,1}
	\end{split}\label{psi1ef}
\end{equation}
by setting $m=0$. Likewise, the $\psi_+\psi_-$ relation includes
\begin{equation}
	\psi^{(a)}_{+,m}\psi^{(b)}_{-,0}=\psi^{(b)}_{-,0}\psi^{(a)}_{+,m},\quad\psi^{(a)}_{+,0}\psi^{(b)}_{-,n}=\psi^{(b)}_{-,n}\psi^{(a)}_{+,0}
\end{equation}
by taking $n=0$ and $m=-2$ respectively. Therefore, $\psi_{\pm,0}$ commute with all the modes of $\psi_{\pm}$. It is worth noting that given a fixed fermionic node $\digamma$, the $\psi^{(\digamma)}_{\pm}$ modes commute with all $\digamma$ modes, and the $e^{(\digamma)}$ (resp. $f^{(\digamma)}$) modes anticommute with the $e^{(\digamma)}$ (resp. $f^{(\digamma)}$) modes. In fact, from \eqref{psi0ef}, it is not hard to see that $\psi^{(a)}_{+,0}\psi^{(a)}_{-,0}$ is central for any node $a$. Write these central elements as $\mathtt{C}^{(a)}=\psi^{(a)}_{+,0}\psi^{(a)}_{-,0}$. Then we can write $\psi^{(a)}_{\pm,0}=\mathtt{C}^{(a)}\left(\psi^{(a)}_{\mp,0}\right)^{-1}$ with a mild assumption that $\left(\psi^{(a)}_{\pm,0}\right)^{-1}$ are also in the algebra. For convenience, we shall rescale them to be 1, that is, $\psi^{(a)}_{+,0}=\left(\psi^{(a)}_{-,0}\right)^{-1}$, in the following discussions.

Like many toroidal algebras, it is instructive to write the $\psi^{(a)}_{\pm}(U)$ currents as
\begin{equation}
	\psi^{(a)}_{\pm}(U)=\psi^{(a)}_{\pm,0}\exp\left(\sum_{n=1}^{\infty}k^{(a)}_{\pm n}U^{\mp n}\right).
\end{equation}
Therefore,
\begin{equation}
	\psi^{(a)}_{\pm,n}=\psi^{(a)}_{\pm,0}\sum_{m=1}^n\frac{1}{m!}\sum_{\substack{r_1,\dots,r_m>0\\r_1+\dots+r_m=n}}k^{(a)}_{\pm r_1}k^{(a)}_{\pm,r_2}\dots k^{(a)}_{\pm,r_m}.
\end{equation}
Similarly, we can write the zero modes as
\begin{equation}
	\psi^{(a)}_{+,0}=\exp\left(-\beta h_1k^{(a)}_0\right)=H_1^{-k_0^{(a)}},\quad\psi^{(a)}_{-,0}=\exp\left(\beta h_1k^{(a)}_0\right)=H_1^{k_0^{(a)}}.
\end{equation}
We shall refer to the modes $k^{(a)}_r$ ($r\in\mathbb{Z}$) as Heisenberg modes. There could also be different conventions to define these modes as discussed in Appendix \ref{kmodes}.

In terms of the Heisenberg modes, we can rewrite the relations involving $\psi_{\pm}$ as
\begin{align}
	&\left[k^{(a)}_0,k^{(b)}_s\right]=0,\quad\left[k^{(a)}_{r\neq0},k^{(b)}_s\right]=\delta_{r+s,0}\frac{1}{r}\left(C^{-r}-C^r\right)H_2^{-rM_{ab}}\left(H_1^{rA_{ab}}-H_1^{-rA_{ab}}\right),\\
	&\left[k^{(a)}_0,e^{(b)}_n\right]=-A_{ab}e^{(b)}_n,\quad\left[k^{(a)}_0,f^{(b)}_n\right]=A_{ab}f^{(b)}_n,\\
	&\left[k^{(a)}_{r\neq0},e^{(b)}_n\right]=\frac{1}{r}C^{-|r|/2}H_2^{-rM_{ab}}\left(H_1^{rA_{ab}}-H_1^{-rA_{ab}}\right)e^{(b)}_{n+r},\\
	&\left[k^{(a)}_{r\neq0},f^{(b)}_n\right]=-\frac{1}{r}C^{|r|/2}H_2^{-rM_{ab}}\left(H_1^{rA_{ab}}-H_1^{-rA_{ab}}\right)f^{(b)}_{n+r}.
\end{align}
Moreover, we have
\begin{equation}
	\left[e^{(a)}_n,f^{(b)}_{-n}\right\}=\delta_{ab}\left(H_1^{k^{(a)}_0}-H_1^{-k^{(a)}_0}\right).
\end{equation}
It would also be helpful to notice that
\begin{equation}
	\left[e^{(a)}_{\pm1},f^{(b)}_0\right\}=\mp\delta_{ab}C^{1/2}H_1^{\mp k^{(a)}_0}k^{(a)}_{\pm1},\quad\left[e^{(a)}_0,f^{(b)}_{\pm1}\right\}=\mp\delta_{ab}C^{-1/2}H_1^{\mp k^{(a)}_0}k^{(a)}_{\pm1}.
\end{equation}

\paragraph{Coproduct} We can also write the coproduct above using $k^{(a)}_r$:
\begin{equation}
	\Delta\left(k^{(a)}_r\right)=\begin{cases}
		C^r\otimes k^{(a)}_r+k^{(a)}_r\otimes1,&r\geq0\\
		k^{(a)}_r\otimes C^r+1\otimes k^{(a)}_r,&r<0
	\end{cases}.
\end{equation}
In particular, $\Delta\left(k^{(a)}_0\right)=1\otimes k^{(a)}_0+k^{(a)}_0\otimes1$.

\paragraph{Grading} Likewise, for the aforementioned grading, we have $\text{deg}\left(k^{(a)}_r\right)=(\bm{0},r)$. In \cite{feigin2015quantum,garbali2021r}, such gradings were useful in the quantum double construction of the universal $R$-matrix for certain toroidal algebra associated to $\mathfrak{gl}_1$. For toroidal BPS algebras associated to any non-chiral quivers here, a naive generalization would be $R=R^{(0)}R^{(1)}R^{(2)}$ with
\begin{equation}
	\begin{split}
		&R^{(0)}=\left(C^{-1}\otimes D^{-1}\right)\left(D^{-1}\otimes C^{-1}\right)\prod_a\left(\psi^{(a)}_{+,0}\otimes\left(D^{(a)}\right)^{-1}\right)\left(\left(D^{(a)}\right)^{-1}\otimes\psi^{(a)}_{+,0}\right),\\
		&R^{(1)}=\text{exp}\left(\sum_{r\geq1}r\sum_ak^{(a)}_r\otimes k^{(a)}_{-r}\right),\quad R^{(2)}=1\otimes1+\sum_{n\in\mathbb{Z}}\sum_ae^{(a)}_n\otimes f^{(a)}_{-n}+\dots,
	\end{split}
\end{equation}
where the ellipsis in $R^{(2)}$ indicates terms with $\text{hdeg}\geq1$, and $\text{pdeg}\left(R^{(2)}\right)$ should be $\bm{0}$. However, whether these naive expressions would work and/or what modifications (such as proper normalizations etc.) are needed would still require further investigations in future.

\subsection{Toric Duality}\label{toricglmn}
Now let us try to construct the transformations of the generators under toric duality. As mentioned in Appendix \ref{gencon}, only fermionic nodes can be dualized. If the node $\digamma$ is dualized, then we just need to add an adjoint loop to $\digamma\pm1$ if $|\digamma\pm1|=0$ or remove the existing adjoint loop on $\digamma\pm1$ if $|\digamma\pm1|=1$. As a result,
\begin{equation}
	\varsigma'_a=\begin{cases}
		-\varsigma_a,&a=\digamma,\digamma+1\\
		\varsigma_a,&\text{otherwise}
	\end{cases},
\end{equation}
where the primed notation stands for the one after performing the duality. Therefore, we have
\begin{equation}
	A'_{ab}=\begin{cases}
		-A_{ab},&(a,b)=(\digamma\pm1,\digamma),(\digamma,\digamma\pm1)\\
		A_{aa}+2A_{a\digamma},&a=b=\digamma\pm1\\
		A_{ab},&\text{otherwise}
	\end{cases}
\end{equation}
and
\begin{equation}
	M'_{ab}=\begin{cases}
		-M_{ab},&a=\digamma-1,\digamma,b=a+1\\
		-M_{ab},&a=\digamma,\digamma+1,b=a-1\\
		M_{ab},&\text{otherwise}
	\end{cases}.
\end{equation}

Analogous to the rational case, the $ke$ and $kf$ commutation relations can be used to express higher $e$, $f$ using lower modes\footnote{Here, by higher (resp. lower) modes, we mean the modes with larger (resp. smaller) absolute values $|n|$.}. The higher modes of $k$ can in turn be obtained using the $ef$ relations. In fact, the relations involving higher modes can also be derived from those with lower modes. Therefore, the toroidal BPS algebras for non-chiral quivers are finitely presented with the relations involving $e_0$, $e_{\pm1}$, $f_0$, $f_{\pm1}$, $k_0$, $k_{\pm1}$ (or equivalently, $\psi_{\pm,0}$, $\psi_{\pm,1}$). Hence, it suffices to find the transformations for these modes\footnote{As pointed out in \cite{Bao:2022jhy}, there is a subtlety for the case $xy=z^2w^2$. For one of the two toric phases, i.e. the one with only fermionic nodes, it seems that the Serre relations can not be fully recovered from the Serre relations for modes with $n=0,\pm1$. However, we may still verify its transformation when using the currents as will be discussed later.}.

We would like to mimic the isomorphisms for the rational case in \cite{Bao:2022jhy}, which was in turn found by virtue of the odd reflections of the underlying affine Lie superalgebras. As all but three of the nodes are unaffected, we would expect the modes to be invariant for $a\neq\digamma,\digamma\pm1$. Therefore, from their relations, we have
\begin{equation}
	C'=C.
\end{equation}

Now, let us first consider the zero modes. For $a=\digamma$, the $k'_0$ modes should be determined only by $k_0$ themselves, possibly with changes of minus signs (such as multiplication by $-1$), while the $e_0$ and $f_0$ modes should get swapped. In the rational case, the $\psi'_0$ mode is a sum of $\psi^{(a)}_0$ and $\psi^{(\digamma)}_0$ for $a=\digamma\pm1$. Here, our ansatz for $\psi_0$ would still be a combination of $\psi^{(a)}_0$ and $\psi^{(\digamma)}_0$, but we expect it to be a multiplication instead of addition as we are dealing with the trigonometric case (and hence addition for $k_0$). On the other hand, for $e'^{(a)}_0$, the ansatz would be a linear combination of $e^{(a)}_0e^{(\digamma)}_0$ and $e^{(\digamma)}_0e^{(a)}_0$ (and likewise for $f'^{(a)}_0$).

By computing the supercommutators $[x,y\}$ with
\begin{equation}
	x=e^{(\digamma)}_0e^{(a)}_0,e^{(a)}_0e^{(\digamma)}_0\text{ and }y=f^{(a)}_0f^{(\digamma)}_0,f^{(\digamma)}_0f^{(a)}_0,
\end{equation}
we find that for $a=\digamma\pm1$,
\begin{equation}
	\begin{split}
		&\psi'^{(a)}_{\pm,0}=\psi^{(a)}_{\pm,0}\psi^{(\digamma)}_{\pm,0},\quad k'^{(a)}_0=k^{(a)}_0+k^{(\digamma)}_0,\\
		&e'^{(a)}_0=e^{(\digamma)}_0e^{(a)}_0-(-1)^{|a|}H_1^{A_{a\digamma}}e^{(a)}_0e^{(\digamma)}_0,\\
		&f'^{(a)}_0=\frac{1}{H_1^{A_{a\digamma}}-H_1^{-A_{a\digamma}}}\left(f^{(a)}_0f^{(\digamma)}_0-(-1)^{|a|}H_1^{-A_{a\digamma}}f^{(\digamma)}_0f^{(a)}_0\right)
	\end{split}
\end{equation}
would verify the corresponding $ef$ relation. Likewise, checking the $ef$ relation for $a=\digamma$, we have
\begin{equation}
	\psi'^{(\digamma)}_{\pm,0}=\psi^{(\digamma)}_{\mp,0},\quad k'^{(\digamma)}_0=-k^{(\digamma)}_0,
\end{equation}
and $e'^{(\digamma)}_0=f^{(\digamma)}_0$, $f'^{(\digamma)}_0=-e^{(\digamma)}_0$. However, to be compatible with the $ee$ and $ff$ relations that contain modes with $n=0,\pm1$, we need to multiply them by some extra factors:
\begin{equation}
	e'^{(\digamma)}_0=\psi^{(\digamma)}_{+,0}f^{(\digamma)}_0=H_1^{-k^{(\digamma)}_0}f^{(\digamma)}_0,\quad f'^{(\digamma)}_0=-\psi^{(\digamma)}_{-,0}e^{(\digamma)}_0=-H_1^{k^{(\digamma)}_0}e^{(\digamma)}_0.
\end{equation}
Notice that they would still recover the transformations of the Chevalley generators under odd reflections in the limit $\beta\rightarrow0$. One may check that these transformations are consistent with all the other relations involving zero modes.

Next, let us consider the modes with $n=\pm1$. By considering the commutator of $k'^{(b\neq\digamma)}_1$ and $e'^{(a)}_0$ with $b=a\pm1$ (which is always possible since there are at least four nodes in the quiver), we find that for $a=\digamma\pm1$,
\begin{equation}
	e'^{(a)}_1=e^{(\digamma)}_0e^{(a)}_1-(-1)^{|a|}H_1^{A_{a\digamma}}e^{(a)}_1e^{(\digamma)}_0.
\end{equation}
Likewise,
\begin{equation}
	f'^{(a)}_1=\frac{1}{H_1^{A_{a\digamma}}-H_1^{-A_{a\digamma}}}\left(f^{(a)}_1f^{(\digamma)}_0-(-1)^{|a|}H_1^{-A_{a\digamma}}f^{(\digamma)}_0f^{(a)}_1\right).
\end{equation}
Again, computing $[x,y\}$ with
\begin{equation}
	x=e^{(\digamma)}_0e^{(a)}_1,e^{(a)}_1e^{(\digamma)}_0\text{ and }y=f^{(a)}_1f^{(\digamma)}_0,f^{(\digamma)}_0f^{(a)}_1,
\end{equation}
we find that
\begin{align}
	&\psi'^{(a)}_{+,1}=\psi^{(\digamma)}_{+,0}\psi^{(a)}_{+,1}-C^{1/2}H_2^{-M_{a\digamma}}\left(H_1^{A_{a\digamma}}f^{(\digamma)}_1e^{(\digamma)}_0+H_1^{-A_{a\digamma}}e^{(\digamma)}_0f^{(\digamma)}_1\right)\psi^{(a)}_{+,0},\\
	&\psi'^{(a)}_{-,1}=\psi^{(\digamma)}_{-,0}\psi^{(a)}_{-,1}-C^{-1/2}H_2^{M_{a\digamma}}\left(H_1^{A_{a\digamma}}e^{(\digamma)}_{-1}f^{(\digamma)}_0+H_1^{-A_{a\digamma}}f^{(\digamma)}_0e^{(\digamma)}_{-1}\right)\psi^{(a)}_{-,0}.
\end{align}
In terms of the Heisenberg modes, we have
\begin{align}
	&k'^{(a)}_1=k^{(a)}_1-C^{1/2}H_2^{-M_{a\digamma}}\left(H_1^{A_{a\digamma}}f^{(\digamma)}_1e^{(\digamma)}_0+H_1^{-A_{a\digamma}}e^{(\digamma)}_0f^{(\digamma)}_1\right)H_1^{k^{(\digamma)}_0},\\
	&k'^{(a)}_{-1}=k^{(a)}_{-1}-C^{-1/2}H_2^{M_{a\digamma}}\left(H_1^{A_{a\digamma}}e^{(\digamma)}_{-1}f^{(\digamma)}_0+H_1^{-A_{a\digamma}}f^{(\digamma)}_0e^{(\digamma)}_{-1}\right)H_1^{-k^{(\digamma)}_0}.
\end{align}
By considering the commutation relations of $k'^{(\digamma\pm1)}_1$ and $e'^{(\digamma)}_0$, we find that
\begin{equation}
	e'^{(\digamma)}_1=CH_2^{-2M_{a\digamma}}H_1^{k^{(\digamma)}_0}f^{(\digamma)}_1,
\end{equation}
where $a$ can either be $\digamma+1$ or $\digamma-1$ as $M_{a\digamma}$ would be the same. Likewise,
\begin{align}
	&f'^{(\digamma)}_1=H_2^{-2M_{a\digamma}}\left(-C^{-1}e^{(\digamma)}_1+C^{-1/2}k^{(\digamma)}_1e^{(\digamma)}_0\right)H_1^{k^{(\digamma)}_0},\\
	&e'^{(\digamma)}_{-1}=H_2^{2M_{a\digamma}}\left(Cf^{(\digamma)}_{-1}-C^{1/2}k^{(\digamma)}_{-1}f^{(\digamma)}_0\right)H_1^{-k^{(\digamma)}_0},\\
	&f'^{(\digamma)}_{-1}=-C^{-1}H_2^{2M_{a\digamma}}H_1^{-k^{(\digamma)}_0}e^{(\digamma)}_{-1}.
\end{align}
Using the $ef$ relations, we get
\begin{equation}
	\psi'^{(\digamma)}_{\pm,1}=-H_2^{\mp2M_{a\digamma}}\left(\psi^{(\digamma)}_{\mp,0}\right)^2\psi^{(\digamma)}_{\pm,1},\quad k'^{(\digamma)}_{\pm1}=-H_2^{\mp2M_{a\digamma}}k^{(\digamma)}_{\pm1}.
\end{equation}
One may check that these transformations are consistent with all the other relations.

From the above discussions, we may also derive the transformations in terms of currents. By applying the $k_{\pm1}$ modes successively, it is not hard to see that
\begin{align}
	&e'^{(a)}(U)=e^{(\digamma)}_0e^{(a)}(U)-(-1)^{|a|}H_1^{A_{a\digamma}}e^{(a)}(U)e^{(\digamma)}_0,\\
	&f'^{(a)}(U)=\frac{1}{H_1^{A_{a\digamma}}-H_1^{-A_{a\digamma}}}\left(f^{(a)}(U)f^{(\digamma)}_0-(-1)^{|a|}H_1^{-A_{a\digamma}}f^{(\digamma)}_0f^{(a)}(U)\right)
\end{align}
for $a=\digamma\pm1$. Then by considering their supercommutator, we find that each term contains some formal delta function with other terms being cancelled. This yields
\begin{equation}
	\begin{split}
		\psi'^{(a)}_{\pm}(U)=&e^{(\digamma)}_0\psi^{(a)}_{\pm}(U)f^{(\digamma)}_0-(-1)^{|a|}H_1^{A_{a\digamma}}e^{(\digamma)}_0f^{(\digamma)}_0\psi^{(a)}_{\pm}(U)\\
		&-H_1^{-A_{a\digamma}}\psi^{(a)}_{\pm}(U)e^{(\digamma)}_0f^{(\digamma)}_0-f^{(\digamma)}_0\psi^{(a)}_{\pm}(U)e^{(\digamma)}_0.
	\end{split}
\end{equation}

It is less straightforward to write down the currents for $\digamma$. Nevertheless, we can write some conjectural expressions by computing a few more higher modes and then verify them using the current relations. The perturbative calculations show that
\begin{align}
	&e'^{(\digamma)}_0(U)=f^{(\digamma)}_{>0}\left(C^{-1}U\right)\overline{\psi}^{(\digamma)}_+\left(C^{-1/2}H_2^{2M_{a\digamma}}U\right)+f^{(\digamma)}_{\leq0}\left(CU\right)\overline{\psi}^{(\digamma)}_-\left(C^{1/2}H_2^{2M_{a\digamma}}U\right),\nonumber\\
	&\\
	&f'^{(\digamma)}_0(U)=-e^{(\digamma)}_{\geq0}\left(CU\right)\overline{\psi}^{(\digamma)}_+\left(C^{1/2}H_2^{2M_{a\digamma}}U\right)-e^{(\digamma)}_{<0}\left(C^{-1}U\right)\overline{\psi}^{(\digamma)}_-\left(C^{-1/2}H_2^{2M_{a\digamma}}U\right),
\end{align}
where
\begin{equation}
	\begin{split}
		&f^{(\digamma)}_{>0}(U)=\sum_{n>0}f^{(\digamma)}_nU^{-n},\quad f^{(\digamma)}_{\leq0}(U)=\sum_{n\leq0}f^{(\digamma)}_nU^{-n},\\ &e^{(\digamma)}_{\geq0}(U)=\sum_{n\geq0}e^{(\digamma)}_nU^{-n},\quad e^{(\digamma)}_{<0}(U)=\sum_{n<0}e^{(\digamma)}_nU^{-n},
	\end{split}
\end{equation}
and
\begin{align}
	\overline{\psi}^{(\digamma)}_+(U)=&\left(\psi^{(\digamma)}_{-,0}\right)^2\left(\psi^{(\digamma)}_{+,0}-\frac{\psi^{(\digamma)}_{+,1}}{U}-\frac{\psi^{(\digamma)}_{+,2}-\left(\psi^{(\digamma)}_{+,1}\right)^2\psi^{(\digamma)}_{-,0}}{U^2}\right.\nonumber\\
	&\qquad\qquad\quad\left.-\frac{\psi^{(\digamma)}_{+,3}+\left(\psi^{(\digamma)}_{+,1}\right)^3\left(\psi^{(\digamma)}_{-,0}\right)^2}{U^3}-\dots\right),\\
	\overline{\psi}^{(\digamma)}_-(U)=&\left(\psi^{(\digamma)}_{+,0}\right)^2\left(\psi^{(\digamma)}_{-,0}-\psi^{(\digamma)}_{-,1}U-\left(\psi^{(\digamma)}_{-,2}-\left(\psi^{(\digamma)}_{-,1}\right)^2\psi^{(\digamma)}_{+,0}\right)U^2\right.\nonumber\\
	&\qquad\qquad\quad\left.-\left(\psi^{(\digamma)}_{-,3}+\left(\psi^{(\digamma)}_{-,1}\right)^3\left(\psi^{(\digamma)}_{+,0}\right)^2\right)U^3-\dots\right).
\end{align}
In fact, we find that the perturbative expressions here coincide with the ``inverse currents'',that is,
\begin{equation}
	\overline{\psi}^{(\digamma)}_{\pm}(U)=\psi^{(\digamma)}_{\pm}(U)^{-1}.
\end{equation}
Then we have
\begin{equation}
	\psi'^{(\digamma)}_{\pm}(U)=\psi^{(\digamma)}_{\pm}\left(H_2^{2M_{a\digamma}}U\right)^{-1}.
\end{equation}
Indeed, one may verify these expressions using the current relations. It is also worth noting that
\begin{equation}
	k'^{(\digamma)}_n=-H_2^{2nM_{a\digamma}}k^{(\digamma)}_n.
\end{equation}

\subsection{Higgsing}\label{higgsingnonchiral}
As reviewed in Appendix \ref{higgsing}, the toric quiver gauge theories have nice features under the Higgs-Kibble mechanism. It is then natural to wonder if their BPS algebras are also connected via blowing up/down the singularities, or more precisely, if there is a subalgebra structure for a higgsed theory from a parent theory.

As the higgsing process always merges the two neighbouring nodes, say $a$ and $a+1$, in the quiver for any toric CY without compact divisors, we expect the generators associated with other nodes (and the central element $C$) to be invariant. Of course, there is a relabelling for $b>a+1$ as the number of nodes is reduced by one after higgsing.

For $x'^{(a)}$ ($x=e,f,\psi,k$), where the primed letters indicate the generators for the higgsed theory, it should be a combination of $x^{(a)}$ and $x^{(a+1)}$. As discussed in Appendix \ref{gencon}, the parity should satisfy $\left|x'^{(a)}\right|=\left|x^{(a)}\right|+\left|x^{(a+1)}\right|$. Therefore, for the zero modes, a natural candidate would be a combination of $e^{(a)}_0e^{(a+1)}_0$ and $e^{(a+1)}_0e^{(a)}_0$ (and likewise for $f$). Similar to the construction for toric duality, we find that
\begin{align}
	&e'^{(a)}_0=e^{(a+1)}_0e^{(a)}_0-(-1)^{|a||a+1|}H_1^{A_{a,a+1}}e^{(a)}_0e^{(a+1)}_0,\\
	&f'^{(a)}_0=\frac{1}{H_1^{A_{a,a+1}}-H_1^{-A_{a,a+1}}}\left(f^{(a)}_0f^{(a+1)}_0-(-1)^{|a||a+1|}H_1^{-A_{a,a+1}}f^{(a+1)}_0f^{(a)}_0\right),\\
	&\psi'^{(a)}_{\pm,0}=\psi^{(a)}_{\pm,0}\psi^{(a+1)}_{\pm,0},\quad k'^{(a)}_0=k^{(a)}_0+k^{(a+1)}_0
\end{align}
would give the expected subalgebra structure for the zero modes. This is precisely the transformation for $a=\digamma\pm1$ in the above discussions of toric duality with $\digamma$ replaced by $a+1$. In fact, in the rational limit $\beta\rightarrow0$, this gives the surjection map of the Chevalley generators of the corresponding affine Lie superalgebras.

However, when we use $k'^{(a-1)}_{\pm1}=k^{(a)}_{\pm1}$ or $k'^{(a+1)}_{\pm1}=k^{(a+2)}_{\pm1}$ to get the higher modes from $e'^{(a)}_0$ (resp. $f'^{(a)}_0$), the expressions are not symmetric in $e^{(a)}$ and $e^{(a+1)}$ (resp. $f^{(a)}$ and $f^{(a+1)}$) any more. Indeed, for instance, $\left[k^{(a-1)}_1,e'^{(a)}_0\right]$ yields
\begin{equation}
	e'^{(a)}_1=e^{(a+1)}_0e^{(a)}_1-(-1)^{|a||a+1|}H_1^{A_{a,a+1}}e^{(a)}_1e^{(a+1)}_0
\end{equation}
while $\left[k^{(a+2)}_1,e'^{(a)}_0\right]$ leads to
\begin{equation}
	e'^{(a)}_1=e^{(a+1)}_1e^{(a)}_0-(-1)^{|a||a+1|}H_1^{A_{a,a+1}}e^{(a)}_0e^{(a+1)}_1.
\end{equation}
They are not equal to each other as can be seen from the $ee$ relation. Explicitly,
\begin{equation}
	e^{(a+1)}_1e^{(a)}_0-(-1)^{|a||a+1|}H_1^{A_{a,a+1}}e^{(a)}_0e^{(a+1)}_1=H_2^{M_{a,a+1}}\left(e^{(a+1)}_0e^{(a)}_1-(-1)^{|a||a+1|}H_1^{A_{a,a+1}}e^{(a)}_1e^{(a+1)}_0\right).
\end{equation}
Due to the non-trivial factor $H_2^{M_{a,a+1}}$, this transformation does not give the subalgebra structure. Nevertheless, when $H_2=1$, the quiver BPS algebras reduce to a one-parameter algebra, and the above two expressions for $e'^{(a)}_1$ would coincide.

Therefore, at least when $h_2=0$, for non-chiral quivers\footnote{For $\mathbb{C}^3/(\mathbb{Z}_2\times\mathbb{Z}_2)$ which can be higgsed to the suspended pinch point, this should also be true. The discussions here do not cover $\mathbb{C}^3$, $\mathbb{C}\times\mathbb{C}^2/\mathbb{Z}_2$ and the conifold although we still expect this to hold.}, the toroidal BPS algebra contains the ones for the higgsed theories as its subalgebras. The surjection for the generators associated with $a$ and $a+1$ are the same as the transformations for $a=\digamma\pm1$ under toric duality with $\digamma$ replaced by $a\mp1$ \footnote{As a result, this gives two transformations, but they should essentially be the same up to a normalization factor.}. Of course, $a+1$ (as well as $a$) can be either bosonic or fermionic. This is also the case for the rational quiver Yangians, where the surjection map is most conveniently expressed in the $J$ presentation. See (4.5) in \cite{Bao:2022jhy} (with conventions therein). It is not clear whether higgsing would still lead to the subalgebra structure for generic $h_2$, and if so, what the surjection map would be. Physically, the two parameters of the algebra are related to the $\Omega$-background that is used to resolve the singular target space of the supersymmetric quantum mechanics. In particular, the scalars in the vector multiplets would also have non-zero VEVs. Therefore, the algebra structure under higgsing could be closely related to the localizations of the Higgs and Coulomb branches \cite{Galakhov:2020vyb}.

\section{Elliptic Algebras and Chiral Quivers}\label{ellipticandchiral}
Now, let us have a discussion on the remaining cases including the elliptic algebras for non-chiral quivers and the algebras for chiral quivers. Unlike the rational and toroidal algebras for non-chiral quivers, it is more difficult to work with modes. This is due to the existence of $q$-Pochhammer symbols in the elliptic case while for chiral quivers, different CYs/quivers would have different ``minimalistic'' presentations. Therefore, we shall mainly consider the more unified current relations.

\subsection{Elliptic Algebras for Non-Chiral Quivers}\label{elliptic}
Given a generalized conifold $xy=z^Mw^N$ with $M+N\geq3$, the elliptic quiver algebra $\mathtt{E}$ has the relations
\begin{align}
	&\psi^{(a)}_{\pm}(U)\psi^{(b)}_{\pm}(V)=\psi^{(b)}_{\pm}(V)\psi^{(a)}_{\pm}(U),\label{ellipticglmnpsipsi1}\\
	&\nonumber\\
	&\psi^{(a)}_{\pm}(U)\psi^{(b)}_{\mp}(V)=\frac{\left(UCV^{-1}H_1^{A_{ab}}H_2^{M_{ab}};q\right)_{\infty}\left(qU^{-1}C^{-1}VH_1^{-A_{ab}}H_2^{-M_{ab}};q\right)_{\infty}}{\left(U^{-1}C^{-1}VH_1^{A_{ab}}H_2^{-M_{ab}};q\right)_{\infty}\left(qUCV^{-1}H_1^{-A_{ab}}H_2^{M_{ab}};q\right)_{\infty}}\nonumber\\
	&\qquad\qquad\qquad\qquad\frac{\left(U^{-1}CVH_1^{A_{ab}}H_2^{-M_{ab}};q\right)_{\infty}\left(qUC^{-1}V^{-1}H_1^{-A_{ab}}H_2^{M_{ab}};q\right)_{\infty}}{\left(UC^{-1}V^{-1}H_1^{A_{ab}}H_2^{M_{ab}};q\right)_{\infty}\left(qU^{-1}CVH_1^{-A_{ab}}H_2^{-M_{ab}};q\right)_{\infty}}\psi^{(b)}_{\mp}(V)\psi^{(a)}_{\pm}(U)\\
	&\psi^{(a)}_{\pm}(U)e^{(b)}(V)=H_1^{A_{ab}}\frac{\left(U^{-1}C^{\mp\frac{1}{2}}VH_1^{-A_{ab}}H_2^{-M_{ab}};q\right)_{\infty}\left(qUC^{\pm\frac{1}{2}}V^{-1}H_1^{A_{ab}}H_2^{M_{ab}};q\right)_{\infty}}{\left(U^{-1}C^{\mp\frac{1}{2}}VH_1^{A_{ab}}H_2^{-M_{ab}};q\right)_{\infty}\left(qUC^{\pm\frac{1}{2}}V^{-1}H_1^{-A_{ab}}H_2^{M_{ab}};q\right)_{\infty}}e^{(b)}(V)\psi^{(a)}_{\pm}(U)\\
	&\psi^{(a)}_{\pm}(U)f^{(b)}(V)=H_1^{-A_{ab}}\frac{\left(U^{-1}C^{\pm\frac{1}{2}}VH_1^{A_{ab}}H_2^{-M_{ab}};q\right)_{\infty}\left(qUC^{\mp\frac{1}{2}}V^{-1}H_1^{-A_{ab}}H_2^{M_{ab}};q\right)_{\infty}}{\left(U^{-1}C^{\pm\frac{1}{2}}VH_1^{-A_{ab}}H_2^{-M_{ab}};q\right)_{\infty}\left(qUC^{\mp\frac{1}{2}}V^{-1}H_1^{A_{ab}}H_2^{M_{ab}};q\right)_{\infty}}f^{(b)}(V)\psi^{(a)}_{\pm}(U)\\
	&e^{(a)}(U)e^{(b)}(V)=(-1)^{|a||b|}H_1^{A_{ab}}\frac{\left(U^{-1}VH_1^{-A_{ab}}H_2^{-M_{ab}};q\right)_{\infty}\left(qUV^{-1}H_1^{A_{ab}}H_2^{M_{ab}};q\right)_{\infty}}{\left(U^{-1}VH_1^{A_{ab}}H_2^{-M_{ab}};q\right)_{\infty}\left(qUV^{-1}H_1^{-A_{ab}}H_2^{M_{ab}};q\right)_{\infty}}e^{(b)}(V)e^{(a)}(U)\\
	&f^{(a)}(U)f^{(b)}(V)=(-1)^{|a||b|}H_1^{-A_{ab}}\frac{\left(U^{-1}VH_1^{A_{ab}}H_2^{-M_{ab}};q\right)_{\infty}\left(qUV^{-1}H_1^{-A_{ab}}H_2^{M_{ab}};q\right)_{\infty}}{\left(U^{-1}VH_1^{-A_{ab}}H_2^{-M_{ab}};q\right)_{\infty}\left(qUV^{-1}H_1^{A_{ab}}H_2^{M_{ab}};q\right)_{\infty}}f^{(b)}(V)f^{(a)}(U)\\
	&\left[e^{(a)}(U),f^{(b)}(V)\right\}=-\delta_{ab}\left(\delta\left(UV^{-1}C^{-1}\right)\psi^{(a)}_+\left(UC^{-1/2}\right)-\delta\left(UV^{-1}C\right)\psi^{(a)}_-\left(VC^{-1/2}\right)\right)\label{ellipticglmnef}.
\end{align}

Similar to the toroidal case, for any fermionic node $\digamma$, we have $\psi^{(\digamma)}_{\pm}(U)e^{(\digamma)}(V)=e^{(\digamma)}(V)\psi^{(\digamma)}_{\pm}(U)$, $e^{(\digamma)}(U)e^{(\digamma)}(V)=-e^{(\digamma)}(V)e^{(\digamma)}(U)$ etc. Moreover, when the central charge is trivial, that is, $C=1$, $\psi^{(a)}_+(U)$ commutes with $\psi^{(b)}_-(V)$.

\subsubsection{More on Mode Expansions}\label{modeselliptic}
Although we would like to work with the currents directly, it would still be helpful to have a look at their mode expansions. There are infinitely many groups of relations as $\alpha$ can be any non-negative integer, but there are finitely many terms in each relation at each order. At order $q^0$, for instance, the $ee$ relations read
\begin{equation}
	e^{(a)}_{m+1,0}e^{(b)}_{n,0}-H_1^{A_{ab}}H_2^{-M_{ab}}e^{(a)}_{m,0}e^{(b)}_{n+1,0}=(-1)^{|a||b|}\left(H_1^{A_{ab}}e^{(b)}_{n,0}e^{(a)}_{m+1,0}-H_2^{-M_{ab}}e^{(b)}_{n+1,0}e^{(a)}_{m,0}\right),
\end{equation}
which coincide with the $ee$ relations for the toroidal algebra. In fact, all the relations at $q^0$ are the same as those in the toroidal case. Therefore, the elliptic subalgebra $\mathtt{E}_0$ at order $q^0$ is isomorphic to the toroidal algebra $\mathtt{T}$. This is expected as the elliptic algebra $\mathtt{E}$ reduces to $\mathtt{T}$ in the limit $q\rightarrow0$.

As another example, let us also write the $\psi e$ relations at order $q^1$ here:
\begin{equation}
	\begin{split}
		\quad\left(H_2^{M_{ab}}U-H_1^{A_{ab}}V\right)&\left(\left(-H_1^{-A_{ab}}H_2^{M_{ab}}UV^{-1}-H_1^{A_{ab}}H_2^{-M_{ab}}VU^{-1}\right)\psi^{(a)}_{\pm,0}\left(C^{\mp1/2}U\right)e^{(b)}_0(V)\right.\\
		&\left.\quad+\psi^{(a)}_{\pm,1}\left(C^{\mp1/2}U\right)e^{(b)}_0(V)+\psi^{(a)}_{\pm,0}\left(C^{\mp1/2}U\right)e^{(b)}_1(V)\right)\\
		=\left(H_1^{A_{ab}}H_2^{M_{ab}}U-V\right)&\left(\left(-H_1^{A_{ab}}H_2^{M_{ab}}UV^{-1}-H_1^{-A_{ab}}H_2^{-M_{ab}}VU^{-1}\right)\psi^{(a)}_{\pm,0}\left(C^{\mp1/2}U\right)e^{(b)}_0(V)\right.\\
		&\left.\quad+\psi^{(a)}_{\pm,1}\left(C^{\mp1/2}U\right)e^{(b)}_0(V)+e^{(b)}_1(V)\psi^{(a)}_{\pm,0}\left(C^{\mp1/2}U\right)\right),
	\end{split}
\end{equation}
from which we can write the corresponding mode relations. The other relations can be obtained in a similar manner. For relations at higher orders of $q$, there would be more terms with larger ranges of modes in the coefficients. In general, at order $q^{\alpha}$, the $\psi_\pm\left(C^{\mp1/2}U\right)e(V)$ relations read
\begin{equation}
	\begin{split}
		&\left(H_2^{M_{ab}}U-H_1^{A_{ab}}V\right)\sum_{\gamma=0}^{\alpha}\sum_{\substack{\alpha_1,\alpha_2\\\alpha_1+\alpha_2=\alpha-\gamma}}K_{\gamma}(A_{ab})\psi^{(a)}_{\pm,\alpha_1}\left(C^{\mp1/2}U\right)e^{(b)}_{\alpha_2}(V)\\
		=&\left(H_1^{A_{ab}}H_2^{M_{ab}}U-V\right)\sum_{\gamma=0}^{\alpha}\sum_{\substack{\alpha_1,\alpha_2\\\alpha_1+\alpha_2=\alpha-\gamma}}K_{\gamma}(-A_{ab})e^{(b)}_{\alpha_2}(V)\psi^{(a)}_{\pm,\alpha_1}\left(C^{\mp1/2}U\right)
	\end{split}\label{psieqorders}
\end{equation}
for some functions $K_{\gamma}$ coming from the expansions of (the product of) the $q$-Pochhammer symbols. Here, we have suppressed the other indices and arguments in $K_{\gamma}$ for brevity. In particular, $K_0=1$. The $e(U)e(V)$ relations have the same coefficients (with an extra sign factor $(-1)^{|a||b|}$) while for the $\psi_{\pm}\left(C^{\mp}U\right)f(V)$ and $f(U)f(V)$ relations, we simply have $A_{ab}\leftrightarrow-A_{ab}$ on both sides. We can then write the mode relations at each order of $q$ from these current relations.

\paragraph{Heisenberg modes} Similar to the discussions on the toroidal algebras above, as well as some elliptic deformed algebras in \cite{Kojima:2011cd}, we may expand the $\psi_{\pm}$ modes as
\begin{equation}
	\psi^{(a)}_+(U)=H_1^{-k_0^{(a)}}\exp\left(\sum_{n\neq0}k^{(a)}_nU^{-n}\right),\quad\psi^{(a)}_-(U)=H_1^{l_0^{(a)}}\exp\left(\sum_{n\neq0}l^{(a)}_{-n}U^n\right).
\end{equation}
For convenience, we shall still refer to the $k$ and $l$ modes as Heisenberg modes. Notice that the sums are now over $\mathbb{Z}\backslash\{0\}$. Moreover,
\begin{align}
	&\psi^{(a)}_{+,n}=H_1^{-k_0^{(a)}}\left(\sum_{m=0}^{\infty}\frac{1}{m!}\sum_{\substack{r_i\neq0\\r_1+\dots+r_m=n}}k_{r_1}k_{r_2}\dots k_{r_m}\right),\\
	&\psi^{(a)}_{-,n}=H_1^{-l_0^{(a)}}\left(\sum_{m=0}^{\infty}\frac{1}{m!}\sum_{\substack{r_i\neq0\\r_1+\dots+r_m=n}}l_{r_1}l_{r_2}\dots l_{r_m}\right).
\end{align}
In particular, $k^{(a)}_0$ and $l^{(a)}_0$ are not equal to $\psi^{(a)}_{\pm,0}$ (or $\psi^{(a)}_{\pm,0,0}$) here. Nevertheless, the Heisenberg modes may still play the role that raises or lowers the $e,f$ modes. More explicitly,
\begin{align}
	&\left[k^{(a)}_r,k^{(b)}_s\right]=\left[l^{(a)}_r,l^{(b)}_s\right]=\left[k^{(a)}_0,l^{(b)}_s\right]=\left[k^{(a)}_r,l^{(b)}_0\right]=0,\\
	&\left[k^{(a)}_{r\neq0},l^{(b)}_s\right]=\delta_{r+s,0}\frac{1}{r}\frac{1}{1-q^r}\left(C^{-r}-C^r\right)H_2^{-rM_{ab}}\left(H_1^{rA_{ab}}-q^rH_1^{-rA_{ab}}\right),\\
	&\left[k^{(a)}_0,e^{(b)}_n\right]=-A_{ab}e^{(b)}_n,\quad\left[k^{(a)}_0,f^{(b)}_n\right]=A_{ab}f^{(b)}_n,\\
	&\left[l^{(a)}_0,e^{(b)}_n\right]=A_{ab}e^{(b)}_n,\quad\left[l^{(a)}_0,f^{(b)}_n\right]=-A_{ab}f^{(b)}_n,\\
	&\left[k^{(a)}_{r\neq0},e^{(b)}_n\right]=\frac{1}{r}\frac{1}{1-q^r}C^{-r/2}H_2^{-rM_{ab}}\left(H_1^{rA_{ab}}-H_1^{-rA_{ab}}\right)e^{(b)}_{n+r},\\
	&\left[k^{(a)}_{r\neq0},f^{(b)}_n\right]=-\frac{1}{r}\frac{1}{1-q^r}C^{r/2}H_2^{-rM_{ab}}\left(H_1^{rA_{ab}}-H_1^{-rA_{ab}}\right)f^{(b)}_{n+r},\\
	&\left[l^{(a)}_{r\neq0},e^{(b)}_n\right]=\frac{1}{r}\frac{1}{1-q^r}C^{-r/2}H_2^{-rM_{ab}}\left(H_1^{rA_{ab}}-H_1^{-rA_{ab}}\right)e^{(b)}_{n-r},\\
	&\left[l^{(a)}_{r\neq0},f^{(b)}_n\right]=-\frac{1}{r}\frac{1}{1-q^r}C^{r/2}H_2^{-rM_{ab}}\left(H_1^{rA_{ab}}-H_1^{-rA_{ab}}\right)f^{(b)}_{n-r}.
\end{align}
However, the $ef$ relations in terms of $k$ and $l$ would be quite different from those of the toroidal cases. This is one of the difficulties when discussing toric duality for elliptic algebras.

\subsubsection{Toric Duality}\label{toricelliptic}
Let us have a brief discussion on toric duality for the elliptic cases. In fact, as discussed in Appendix \ref{serre}, the dressed currents $E^{(a)}(u)$, $F^{(a)}(u)$ and $\Psi^{(a)}_{\pm}(u)$ introduced therein have the same relations as those of the toroidal cases. Therefore, the previous transformations should also apply to the elliptic cases using the dressed currents (with products replaced by correlators or normal orderings). Moreover, by comparing these relations with the ones using the ``bare'' generators at each order $q^{\alpha}$, we may write the correlators $\langle XY\rangle_{\alpha}$ in the expansion of $q$. For instance, from \eqref{psieqorders}, we have
\begin{align}
	&\left\langle \Psi^{(a)}_{\pm}\left(C^{\mp1/2}U\right)E^{(b)}\left(V\right)\right\rangle_{\alpha}=\sum_{\gamma=0}^{\alpha}\sum_{\substack{\alpha_1,\alpha_2\\\alpha_1+\alpha_2=\alpha-\gamma}}K_{\gamma}(A_{ab})\psi^{(a)}_{\pm,\alpha_1}\left(C^{\mp1/2}U\right)e^{(b)}_{\alpha_2}(V),\\
	&\left\langle E^{(b)}\left(V\right)\Psi^{(a)}_{\pm}\left(C^{\mp1/2}U\right)\right\rangle_{\alpha}=\sum_{\gamma=0}^{\alpha}\sum_{\substack{\alpha_1,\alpha_2\\\alpha_1+\alpha_2=\alpha-\gamma}}K_{\gamma}(-A_{ab})e^{(b)}_{\alpha_2}(V)\psi^{(a)}_{\pm,\alpha_1}\left(C^{\mp1/2}U\right).
\end{align}
Nevertheless, let us still take a look at the original bare generators $e$, $f$, $\psi_{\pm}$ directly in the followings for completeness.

Suppose that the node $\digamma$ is dualized. Then the currents associated to $a\neq\digamma,\digamma\pm1$ (and hence $C$) should remain invariant. For $a=\digamma\pm1$, we expect the currents to have a combination of $a$ and $\digamma$ currents/modes similar to the ones in the toroidal cases. Let us recall that for the toroidal algebras, we have
\begin{equation}
	e'^{(a)}(U)=\left[e^{(\digamma)}_0,e^{(a)}(U)\right\}_{H_1^{A_{a\digamma}}},
\end{equation}
where the deformed bracket is given by $[x,y\}_{\mathfrak{q}}=xy-(-1)^{|x||y|}\mathfrak{q}yx$. Likewise, for the rational algebras, we have
\begin{equation}
	e'^{(a)}(U)=\left[e^{(\digamma)}_0,e^{(a)}(U)\right\}.
\end{equation}
As a result, each transformation is determined by its corresponding version of the bracket. Moreover, these are preciously the brackets that appear in their Serre relations. Therefore, we propose that the elliptic version of the bracket is used here:
\begin{equation}
	e'^{(a)}(U)=\left.\left[e^{(\digamma)}(V),e^{(a)}(U)\right\}_{\chi}\right\vert_{V^0},
\end{equation}
where $\chi$ represents the elliptic deformed bracket as in Appendix \ref{serre} and $V^0$ indicates that we only take the terms of order $V^0$. More explicitly, using the $q$-binomial theorem, we have
\begin{equation}
	\begin{split}
		&e'^{(a)}(U)=\sum_{n=0}^{\infty}\frac{\left(H_1^{2A_{a\digamma}};q\right)_n}{(q;q)_n}\left(qH_1^{-A_{a\digamma}}H_2^{M_{a\digamma}}U\right)^n\\
		&\qquad\qquad\qquad\left(e^{(\digamma)}_{-n}e^{(a)}(U)-(-1)^{|a|}H_1^{A_{a\digamma}}H_2^{-2nM_{a\digamma}}U^{-2n}e^{(a)}(U)e^{(\digamma)}_n\right).
	\end{split}
\end{equation}
Likewise,
\begin{equation}
	\begin{split}
		&f'^{(a)}(U)=\sum_{n=0}^{\infty}\frac{\left(H_1^{-2A_{a\digamma}};q\right)_n}{(q;q)_n}\frac{\left(qH_1^{A_{a\digamma}}H_2^{-M_{a\digamma}}U^{-1}\right)^n}{H_1^{A_{a\digamma}}-H_1^{-A_{a\digamma}}}\\
		&\qquad\qquad\qquad\left(f^{(a)}(U)f^{(\digamma)}_n-(-1)^{|a|}H_1^{-A_{a\digamma}}H_2^{2nM_{a\digamma}}U^{2n}f^{(\digamma)}_{-n}f^{(a)}(U)\right).
	\end{split}
\end{equation}

For the node $\digamma$, we expect that $\psi'_{\pm}$ are still given by the inverse currents, that is,
\begin{equation}
	\psi'^{(\digamma)}_{\pm}(U)=\psi^{(\digamma)}_{\pm}\left(H_2^{2M_{a\digamma}}U\right)^{-1}.
\end{equation}
Analogously, it is natural to conjecture that $e'^{(\digamma)}$ and $f'^{(\digamma)}$ would have the same forms as in the toroidal algebras. In other words
$e'=f_{>0}\psi_+^{-1}+f_{\leq0}\psi_-^{-1}$, $f'=-e_{\geq0}\psi_+^{-1}-e_{<0}\psi_-^{-1}$, where we have omitted the different arguments in different factors for brevity.

Indeed, the inverse currents are consistent with the relations under toric duality. For instance, the $e'^{(a)}e'^{(\digamma)}$ relation contains
\begin{equation}
	\begin{split}
		&e^{(a)}(U)\mathcal{E}^{(\digamma)}\mathcal{F}^{(\digamma)}_{\pm}\psi^{(\digamma)}_{\pm}\left(C^{\mp1/2}H_2^{2M_{a\digamma}}V\right)^{-1}\\
		=&(-1)^{|a|}U^{-1}VH_2^{M_{a\digamma}}\frac{\left(UV^{-1}H_1^{-A_{a\digamma}}H_2^{-M_{a\digamma}};q\right)_{\infty}}{\left(U^{-1}VH_1^{-A_{a\digamma}}H_2^{M_{a\digamma}};q\right)_{\infty}}\frac{\left(qU^{-1}VH_1^{A_{a\digamma}}H_2^{M_{a\digamma}};q\right)_{\infty}}{\left(qUV^{-1}H_1^{A_{a\digamma}}H_2^{-M_{a\digamma}};q\right)_{\infty}}\\
		&\mathcal{F}^{(\digamma)}_{\pm}\psi^{(\digamma)}_{\pm}\left(C^{\mp1/2}H_2^{2M_{a\digamma}}V\right)^{-1}e^{(a)}(U)\mathcal{E}^{(\digamma)}+\dots,
	\end{split}
\end{equation}
where $\mathcal{E}^{(\digamma)}$ (resp. $\mathcal{F}^{(\digamma)}_{\pm}$) sketchily indicates the factors containing only $e^{(\digamma)}$ (resp. $f^{(\digamma)}$) modes. The ellipsis stands for the extra terms coming from exchanging these factors which should be cancelled in the whole expression. Recall that $A'_{a\digamma}=-A_{a\digamma}$ and $M'_{a\digamma}=-M_{a\digamma}$. As we can see, this recovers the correct coefficient for the $e'^{(a)}e'^{(\digamma)}$ relation.

\paragraph{Higgsing} Similar to the rational and toroidal cases, the surjection (if it exists) induced from higgsing should leave the central element $C$ and all but two (say, $a$ and $a+1$) currents invariant (with a possible relabelling of nodes). However, due to the complication at higher orders of $q$, it is more difficult to write the currents associated to $a'$ in terms of those for $a$ and $a+1$. Nevertheless, we may still conjecture that higgsing would also give subalgebras in the elliptic case, at least in certain one-parameter degeneracy.

\subsection{Comments on Chiral Quivers}\label{chiral}
\begin{figure}[h]
	\centering
	\includegraphics{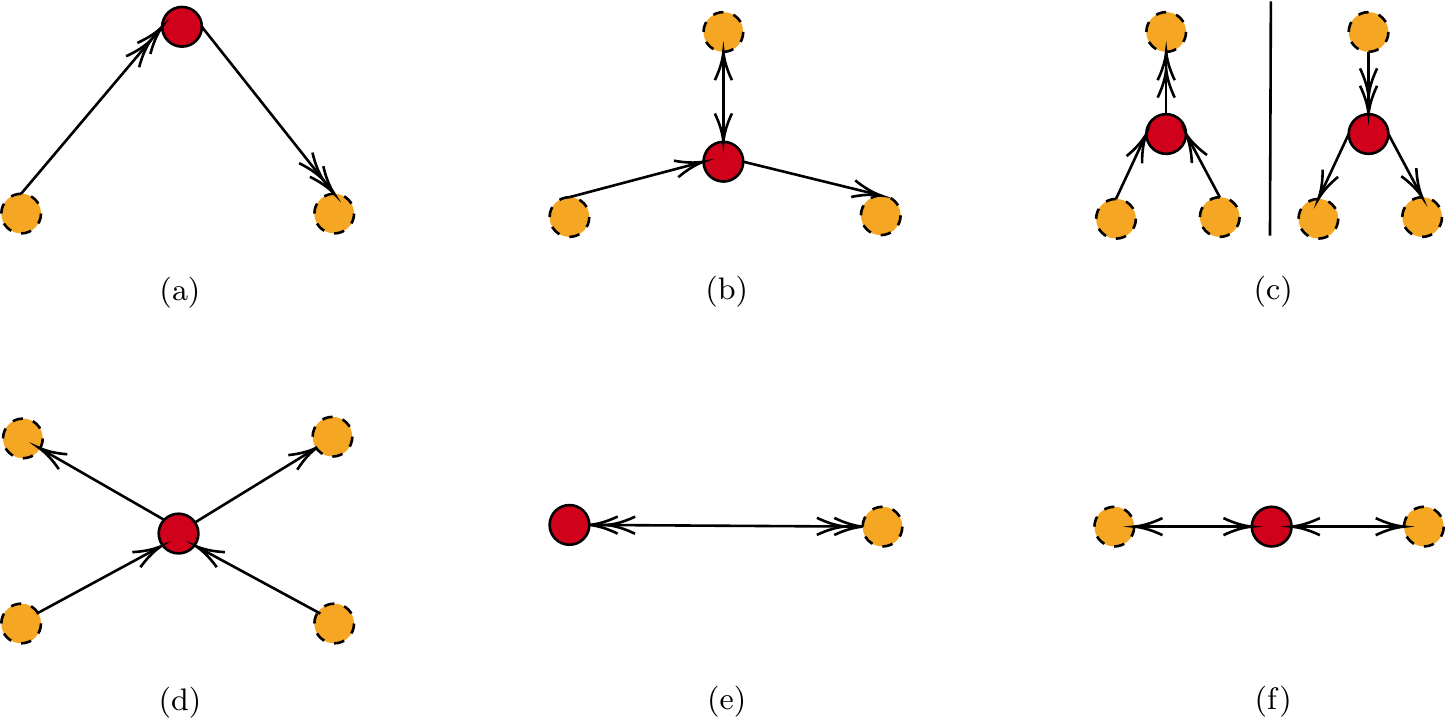}
	\caption{The six possible configurations for the dualized node in the quiver. The node to be dualized is coloured red. The dashed nodes indicate that they can be connected to the remaining part of the quiver.}\label{sixcases}
\end{figure}

We shall now make some comments on the cases for chiral quivers. As mentioned in Appendix \ref{toric}, only nodes with two arrows in and two arrows out will be dualized. Then all the possible cases are listed in Figure \ref{sixcases}. However, as we are now going to discuss, we will only consider the cases (a), (c) and (d) here.

\begin{figure}[h]
	\includegraphics{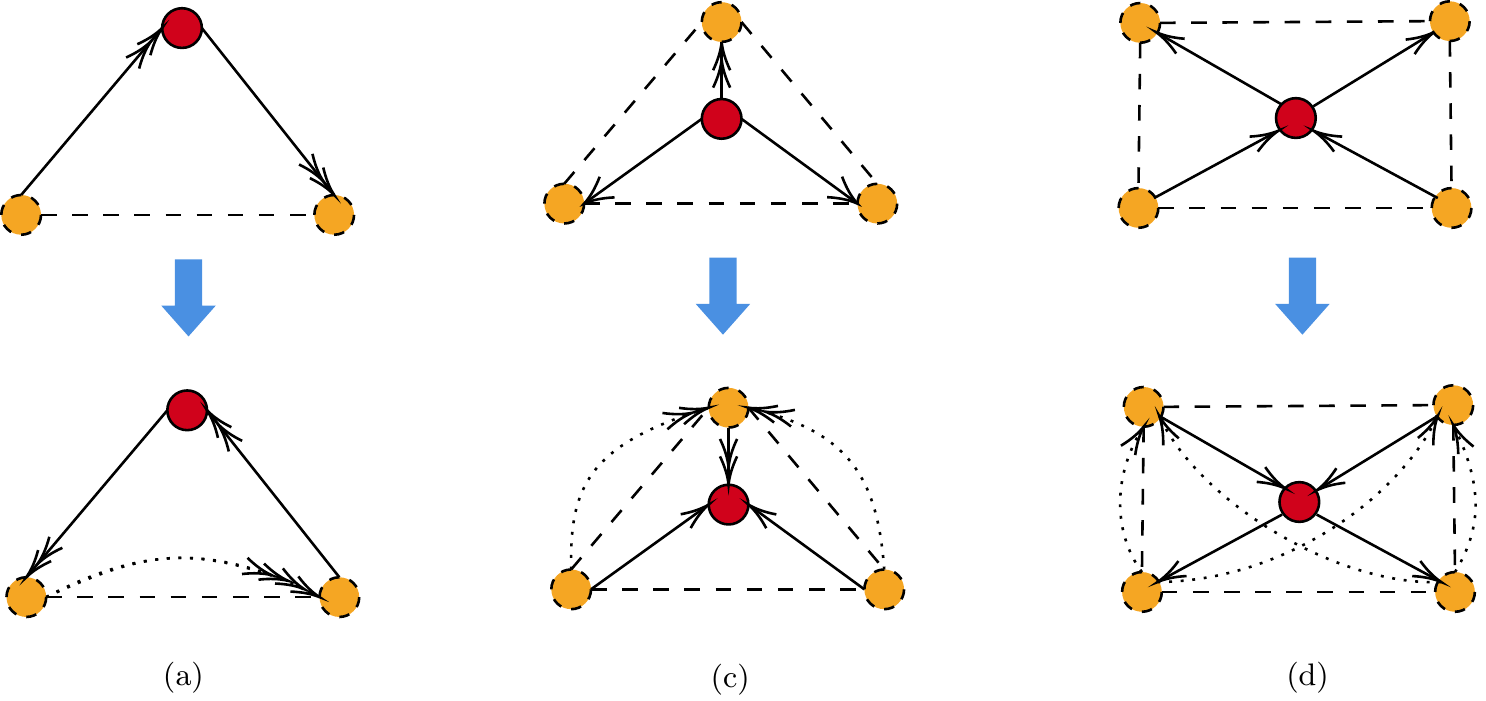}
	\caption{How the arrows and hence the $\zeta$ factors would change under toric duality for (a), (c), (d). The two types of dashed lines indicate the arrows connecting the orange nodes (before possible cancellations).}\label{threecases}
\end{figure}

For the last two cases, (e) and (f), the quivers would remain the same after dualizing the red node (assuming that all the arrows added get integrated out). Therefore, their quiver BPS algebras are trivially invariant, and we only need to focus on the remaining four cases. Moreover, for toric CYs with compact divisors, as the quiver nodes do not have adjoints (at least for all the known examples to our best knowledge), all the $e$ and $f$ modes/currents are fermionic. In particular, this means that (b) should be excluded as the node with two arrows (one in and one out) connected to the dualized node will get an adjoint loop that cannot be integrated out under duality. Indeed, as far as we know, including the examples classified in \cite{Hanany:2012hi,Bao:2020kji,Franco:2017jeo}, there is no case (b) appearing. For the remaining three cases, their quivers under toric duality are illustrated in Figure \ref{threecases}.

As a preliminary attempt of constructing the transformations, let us consider certain expressions similar to the cases for non-chiral quivers. Of course, the central element $C$ and the currents associated to the nodes that are not connected to the dualized node should be invariant.

Suppose that the node $\digamma$ is dualized. As before, we expect $e'^{(\digamma)}(u)$ to be a combination of $\mathcal{F}^{(\digamma)}_{\pm}(u)\psi^{(\digamma)}_{\pm}(-u\mp c/2)^{-1}$ for some $\mathcal{F}^{(\digamma)}_{\pm}(u)$ (and likewise for $f'^{(\digamma)}(u)$). For simplicity, let us just take $e'^{(\digamma)}(u)=f^{(\digamma)}(u)\psi^{(\digamma)}_+(-u-c/2)^{-1}$ as an illustration. Indeed, as $\digamma$ has all its arrow(s) connected to $a$ being reversed, $e^{(a)}(u)f^{(\digamma)}(u)\psi^{(\digamma)}_+(-u-c/2)^{-1}$ would give the required prefactor from the $e^{(a)}\left(\psi^{(\digamma)}\right)^{-1}$ relation while $e^{(a)}f^{(\digamma)}$ would be responsible for the minus sign.

For the nodes connected to $\digamma$, since they would remain fermionic after toric duality, we cannot multiply them by $e^{(\digamma)}$ or $f^{(\digamma)}$ as in the non-chiral quiver cases. Let us first consider the cases (a) and (c). Suppose that we take
\begin{equation}
	e'^{(a)}(u)=\begin{cases}
		e^{(a)}(-u)\psi^{(\digamma)}_+(-u-c/2),&a\twoheadrightarrow\digamma\text{ or }\digamma\twoheadrightarrow a\\
		e^{(a)}(-u),&a\rightarrow\digamma\text{ or }\digamma\rightarrow a
	\end{cases},
\end{equation}
where $a\rightarrow b$ and $a\twoheadrightarrow b$ indicate the number of arrows from $a$ to $b$. Then we have
\begin{equation}
	\begin{split}
		e'^{(a)}(u)e'^{(\digamma)}(v)&=e'^{(a)}(u)f^{(\digamma)}(v)\psi^{(\digamma)}_+(-v-c/2)^{-1}\\
		&=-\phi^{a\Leftarrow\digamma}(-u,-v)^{-1}f^{(\digamma)}(v)\psi^{(\digamma)}_+(-v-c/2)^{-1}e'^{(a)}(u)\\
		&=-\left(UV\right)^{-\frac{\mathfrak{t}}{2}\chi_{a\digamma}}\phi^{a\Leftarrow\digamma}(v-u)^{-1}f^{(\digamma)}(v)\psi^{(\digamma)}_+(-v-c/2)^{-1}e'^{(a)}(u),
	\end{split}
\end{equation}
which recovers the correct numbers of $\zeta$ in the relations as $\chi'_{a\digamma}=-\chi_{a\digamma}$. For instance, if $a\twoheadrightarrow\digamma$ in the original quiver, then we have $\digamma\twoheadrightarrow a$ after toric duality, and
\begin{equation}
	\phi^{a\Leftarrow\digamma}(v-u)^{-1}=\frac{1}{ \zeta\left(h_{a\digamma}^1-u+v\right)\zeta\left(h_{a\digamma}^2-u+v\right)}.\label{zetaex}
\end{equation}
One may also check that the other $e'e'$ relations would also give the correct numbers of the $\zeta$ factors. For the case (d), we may choose
\begin{equation}
	e'^{(a)}(u)=\begin{cases}
		e^{(a)}(-u)\psi^{(\digamma)}_+(-u-c/2),&a\rightarrow\digamma\\
		e^{(a)}(-u),&\digamma\rightarrow a
	\end{cases}.
\end{equation}

However, only checking the numbers of $\zeta$ in the relations is not sufficient, and astute readers may have already found the following problems:
\begin{itemize}
	\item Recall that for the rational quiver Yangians, the equalities in the current relations are up to some $u^mv^n$ terms. The transformations in terms of the currents may not incorporate these terms in general. Whether the transformations in terms of currents would work or how corrections should be made would probably require more detailed delibrations on the relations of modes, which can be much more intricate.
	
	\item From the transformations of $e$ and $f$, we may obtain $\psi'_{\pm}$ from the $ef$ relations. However, unlike the toroidal and elliptic algebras for non-chiral quivers discussed above, there would be terms that do not have formal delta functions. Although we could still in principle put them at the right orders of $U$, $V$ (and $q$) in the mode expansions of $\psi'_{\pm}$, there could be ambiguities in this process. This subtlety should also be related to the ambiguities of multiplying $\psi^{(\digamma)}_{\pm}$ in the above transformations.
	
	\item Most importantly, when we check the $\zeta$ factors above, we have not taken the correct charge assignment for the dual algebra into account. For instance, $h_{a\digamma}^i$ in \eqref{zetaex} may not be the right charges for the arrows in the dualized quiver. In fact, by checking some examples, it is not hard to find that even if \eqref{zetaex} gives the correct charges, the arrows connecting the orange nodes do not have the required charges after the transformations. In fact, such transformations would only work when the two parameters $h_{1,2}$ are both zero. One may consider possible shifts of the spectral parameters, such as $e^{(a)}(-u+\epsilon_1)\psi^{(\digamma)}(-u-c/2+\epsilon_2)$ etc., in the above transformations. However, it would yield a set of homogeneous equations only with the trivial solution as there are more independent constraints than variables.
\end{itemize}
Therefore, the transformations for the dual algebras require a more careful construction. Finding such maps may require more sophisticated ways, and we leave it to future work.

Likewise, for higgsing, simple multiplications of the currents for the merged nodes would only give subalgebra structure in the trivial case with vanishing parameters. Moreover, given a chiral quiver, it can be higgsed to either chiral or non-chiral quivers. There can also be more than one pair of nodes to be merged. Although we still expect such surjection maps under higgsing (at least for one-parameter degeneracies), it could be very different from the above cases involving only non-chiral quivers.

\subsubsection{More on Mode Expansions}\label{modeschiral}
Similar to the discussions for non-chiral quivers, we may also take the mode expansions as
\begin{equation}
	\psi^{(a)}_+(U)=\exp\left(\sum_{n\in\mathbb{Z}}k^{(a)}_nU^{-n}\right),\quad\psi^{(a)}_-(U)=\exp\left(\sum_{n\in\mathbb{Z}}l^{(a)}_{-n}U^n\right).
\end{equation}
We shall still refer to $k$ and $l$ as Heisenberg modes. Notice that the conventions when writing $k_0$ and $l_0$ are slightly different from before, and the sums are over $\mathbb{Z}$.

Consider two nodes $a$ and $b$ in any chiral quiver. Suppose that there are $|a\rightarrow b|=r$ and $|b\rightarrow a|=s$. Then
\begin{align}
	&\left[k^{(a)}_0,l^{(b)}_0\right]=\log\left(C^{-r-s}\right)=-(r+s)\beta c,\\
	&\left[k^{(a)}_0,k^{(b)}_0\right]=-\left[l^{(a)}_0,l^{(b)}_0\right]=\log\left(C^{r-s}\right)=(r-s)\beta c,\\
	&\left[k^{(a)}_{m\neq0},k^{(b)}_n\right]=\left[l^{(a)}_{m\neq0},l^{(b)}_n\right]=\left[k^{(a)}_0,l^{(b)}_{n\neq0}\right]=\left[k^{(a)}_{m\neq0},l^{(b)}_0\right]=0.
\end{align}
Moreover, we have
\begin{equation}
	\left[k^{(a)}_0,e^{(b)}(V)\right]=\left[l^{(a)}_0,e^{(b)}(V)\right]=\begin{cases}
		\log\left(\mathcal{H}_{ab}V^{-(r-s)}\right)e^{(b)}(V),&r>s\\
		\log\left(-\mathcal{H}_{ab}V^{-(r-s)}\right)e^{(b)}(V),&r<s\\
		\log\left((-1)^r\mathcal{H}_{ab}\right)e^{(b)}(V),&r=s\\
	\end{cases},
\end{equation}
\begin{equation}
	\left[k^{(a)}_0,f^{(b)}(V)\right]=\left[l^{(a)}_0,f^{(b)}(V)\right]=\begin{cases}
		-\log\left(\mathcal{H}_{ab}V^{-(r-s)}\right)f^{(b)}(V),&r>s\\
		-\log\left(-\mathcal{H}_{ab}V^{-(r-s)}\right)f^{(b)}(V),&r<s\\
		-\log\left((-1)^r\mathcal{H}_{ab}\right)f^{(b)}(V),&r=s\\
	\end{cases},
\end{equation}
where $\mathcal{H}_{ab}=\prod\limits_{i=1}^rH_{ab,i}^{1/2}\prod\limits_{j=1}^sH_{ba,j}^{1/2}$. It would be more useful to write them as
\begin{align}
	&\text{e}^{\pm\frac{1}{r-s}k^{(a)}_0}e^{(b)}_n\text{e}^{\mp\frac{1}{r-s}k^{(a)}_0}=\text{sgn}(r,s){H}_{ab}^{\pm\frac{1}{r-s}}e^{(b)}_{n\mp1}\quad(r\neq s),\label{k01}\\
	&\text{e}^{\pm\frac{1}{r-s}k^{(a)}_0}f^{(b)}_n\text{e}^{\mp\frac{1}{r-s}k^{(a)}_0}=\text{sgn}(r,s){H}_{ab}^{\mp\frac{1}{r-s}}f^{(b)}_{n\pm1}\quad(r\neq s),\\
	&\text{e}^{k^{(a)}_0}e^{(b)}_n\text{e}^{-k^{(a)}_0}=\text{sgn}(r,s)\mathcal{H}_{ab}e^{(b)}_n\quad(r=s),\\
	&\text{e}^{k^{(a)}_0}f^{(b)}_n\text{e}^{-k^{(a)}_0}=\text{sgn}(r,s)\mathcal{H}_{ab}f^{(b)}_n\quad(r=s),\label{k04}\\
\end{align}
and likewise for $l^{(a)}_0$, where we have defined
\begin{equation}
	\text{sgn}(r,s)=\begin{cases}
		1,&r>s\\
		(-1)^r,&r=s\\
		-1,&r<s
	\end{cases}.
\end{equation}

The remaining relations would be different for the toroidal and the elliptic cases. For the toroidal algebras, we have
\begin{align}
	&\left[k^{(a)}_m,e^{(b)}_n\right]=\frac{1}{m}C^{-m/2}\left(\sum_jH_{ba,j}^m-\sum_iH_{ab,i}^{-m}\right)e^{(b)}_{n+m}\qquad(m>0),\\
	&\left[k^{(a)}_m,f^{(b)}_n\right]=-\frac{1}{m}C^{m/2}\left(\sum_jH_{ba,j}^m-\sum_iH_{ab,i}^{-m}\right)f^{(b)}_{n+m}\qquad(m>0),\\
	&\left[l^{(a)}_{-m},e^{(b)}_n\right]=\frac{1}{m}C^{m/2}\left(\sum_jH_{ba,j}^m-\sum_iH_{ab,i}^{-m}\right)e^{(b)}_{n+m}\qquad(m>0),\\
	&\left[l^{(a)}_{-m},f^{(b)}_n\right]=-\frac{1}{m}C^{-m/2}\left(\sum_jH_{ba,j}^m-\sum_iH_{ab,i}^{-m}\right)f^{(b)}_{n+m}\qquad(m>0),\\
	&\left[k^{(a)}_m,e^{(b)}_n\right]=\left[k^{(a)}_m,f^{(b)}_n\right]=\left[l^{(a)}_{-m},e^{(b)}_n\right]=\left[l^{(a)}_{-m},f^{(b)}_n\right]=0\qquad(m<0),\\
	&\left[k^{(a)}_m,l^{(b)}_n\right]=\delta_{m+n,0}\frac{1}{m}\left(C^{-m}-C^m\right)\left(\delta_{m>0}\sum_jH_{ba,j}^m+\delta_{m<0}\sum_iH_{ab,i}^{-m}\right)\qquad(m\neq0),
\end{align}
where $\delta_{\mathtt{cond}}$ is 1 when the condition $\mathtt{cond}$ is satisfied and 0 otherwise. Notice that we would only raise the $e,f$ modes using the non-zero Heisenberg modes. If we take $\mathfrak{t}=1$ in the balancing factor $(UV)^{\frac{\mathfrak{t}}{2}\chi_{ab}}$ for the toroidal algebras, then only $k_m$ and $l_{-m}$ with $m<0$ would lower the $e,f$ modes while the other non-zero Heisenberg modes would commute with them. This would also make certain changes in \eqref{k01}$\sim$\eqref{k04}.

For the elliptic algebras, we have
\begin{align}
	&\left[k^{(a)}_m,e^{(b)}_n\right]=\frac{1}{m}\frac{1}{1-q^m}C^{-m/2}\left(\sum_jH_{ba,j}^m-\sum_iH_{ab,i}^{-m}\right)e^{(b)}_{n+m},\\
	&\left[k^{(a)}_m,f^{(b)}_n\right]=-\frac{1}{m}\frac{1}{1-q^m}C^{m/2}\left(\sum_jH_{ba,j}^m-\sum_iH_{ab,i}^{-m}\right)f^{(b)}_{n+m},\\
	&\left[l^{(a)}_{-m},e^{(b)}_n\right]=\frac{1}{m}\frac{1}{1-q^m}C^{m/2}\left(\sum_jH_{ba,j}^m-\sum_iH_{ab,i}^{-m}\right)e^{(b)}_{n+m},\\
	&\left[l^{(a)}_{-m},f^{(b)}_n\right]=-\frac{1}{m}\frac{1}{1-q^m}C^{-m/2}\left(\sum_jH_{ba,j}^m-\sum_iH_{ab,i}^{-m}\right)f^{(b)}_{n+m},\\
	&\left[k^{(a)}_m,l^{(b)}_n\right]=\delta_{m+n,0}\frac{1}{m}\frac{1}{1-q^m}\left(C^{-m}-C^m\right)\left(\sum_iH_{ab,i}^{-m}-\sum_jH_{ba,j}^m\right),
\end{align}
where $m\neq0$. If we take $\mathfrak{t}=1$ in the balancing factor $(UV)^{\frac{\mathfrak{t}}{2}\chi_{ab}}$, then $1/(1-q^m)$ would be changed to $1/\left(q^{-m}-1\right)$.

\subsubsection{Free Field Realizations}\label{freefields}
Let us now discuss the free field realizations of the toroidal and elliptic quiver BPS algebras. From the discussions in Appendix \ref{serre}, it suffices to consider the toroidal case. For non-chiral quivers, the level $(1,0)$ representation (when $c=h_1$) was given in \cite{bezerra2021quantum} with notations and conventions therein. Therefore, we shall only mention the cases for chiral quivers here.

Let us rewrite the balancing factor in \eqref{balancing} as $\left(\prod H_{ab,i}\right)^{1/2}\left(\prod H_{ba,i}\right)^{-1/2}(UV)^{\chi_{ab}/2}$ for convenience. We have essentially made two changes here. First, we use the convention $\mathfrak{t}=1$ instead of $-1$. Moreover, the extra factors with $H_{ab,i}$ (and $H_{ba,i}$) are included so as to remove the half integer powers of these parameters (just like what \eqref{balancing} does for the spectral parameters). Of course, these extra factors can always be re-absorbed into the OPEs of the free fields that will be introduced below.

In the remaining part of this section, we shall write $\mathfrak{q}=C$. It would also be convenient to use the standard notation $[n]_{\mathfrak{q}}=\frac{\mathfrak{q}^n-\mathfrak{q}^{-n}}{\mathfrak{q}-\mathfrak{q}^{-1}}$. Let us write the OPE of two vertex operators as
\begin{equation}
	\mathcal{V}_1(Z)\mathcal{V}_2(W)=\langle\mathcal{V}_1(Z)\mathcal{V}_2(W)\rangle(\mathcal{V}_1(Z)\mathcal{V}_2(W)),
\end{equation}
where we have used $(\dots)$ to denote the normal ordering and $\langle\dots\rangle$ is the contraction. In particular, $(\mathcal{V}_1(Z)\mathcal{V}_2(W))=(\mathcal{V}_2(W)\mathcal{V}_1(Z))$. Notice that here, the contraction $\langle\mathcal{V}_1(z)\mathcal{V}_2(w)\rangle$ which is a rational function should be understood as a Laurent series that converges in the region $|Z|\gg|W|$. Therefore, it would be helpful to recall that for any rational function $F(Z)$, we have
\begin{equation}
	\left[F(Z)\right]_{|Z|\gg1}-\left[F(Z)\right]_{|Z|\ll1}=-\sum_i\delta\left(\frac{Z}{\mathtt{r}_i}\right)\text{Res}_{\mathtt{r}_i}\frac{F(Z)}{Z},
\end{equation}
where $[\dots]_{\mathcal{A}}$ denotes the Laurent expansion in the region $\mathcal{A}$ and the sum is over all the poles $\mathtt{r}_i$ of $F$ other than $0$ and $\infty$. As we are actually considering the $\mathfrak{q}$-deformed algebras, we shall also use the difference operator $\partial$ such that
\begin{equation}
	\partial \mathcal{V}(Z)=\frac{\mathcal{V}\left(\mathfrak{q}Z\right)-\mathcal{V}\left(\mathfrak{q}^{-1}Z\right)}{\left(\mathfrak{q}-\mathfrak{q}^{-1}\right)Z}.
\end{equation}

Let us introduce the generators satisfying
\begin{equation}
	\begin{split}
		&\left[x^{(a)}_r,x^{(b)}_s\right]=\delta_{r+s,0}\frac{[r]_{\mathfrak{q}}^2}{r}\sum_i\mathfrak{q}H_{ba,i},\\
		&\left[y^{(b)}_r,y^{(a)}_s\right]=\delta_{r+s,0}\frac{[r]_{\mathfrak{q}}^2}{r}\sum_i\mathfrak{q}H_{ab,i},\\
		&\left[\gamma^{(a)}_r,\gamma^{(b)}_s\right]=\delta_{ab}\delta_{r+s,0}\frac{[r]_{\mathfrak{q}}^2}{r},
	\end{split}
\end{equation}
with the other commutators vanishing. Consider the currents
\begin{equation}
	\begin{split}
		&X^{(a)}(U)=\log(U)+x^{(a)}_-\left(\mathfrak{q}^{-1}U\right)-x^{(a)}_+(U)+x^{(a)}_0\log(U)+\alpha_x^{(a)},\\
		&Y^{(a)}(U)=\log(U)+y^{(a)}_-(\mathfrak{q}^{-1}U)-y^{(a)}_+\left(U\right)-y^{(a)}_0\log(V)-\alpha_y^{(a)},\\
		&\Gamma^{(a)}_{\pm}(U)=\pm\gamma^{(a)}_-\left(U\right)\mp\gamma^{(a)}_+(U)\pm\widetilde{\gamma}^{(a)}\pm\gamma^{(a)}_0\log(U),\\
	\end{split}
\end{equation}
where
\begin{equation}
	x^{(a)}_{\pm}(U)=\sum_{r>0}\frac{x^{(a)}_r}{[r]_{\mathfrak{q}}}U^{\mp r}
\end{equation}
and likewise for $y^{(a)}_{\pm}(U)$, $\gamma^{(a)}_{\pm}(U)$. We have also introduced the elements $\alpha_{x,y}^{(a)}$, $\widetilde{\gamma}^{(a)}$ such that
\begin{equation}
	\begin{split}
		&\left\langle\exp\left(\alpha_x^{(a)}\right)U^{x^{(a)}_0}\right\rangle=U^{|a\rightarrow b|},\quad\left\langle\exp\left(\alpha_y^{(a)}\right)U^{y^{(a)}_0}\right\rangle=U^{-|a\rightarrow b|},\\
		&\left(\exp\left(\alpha_i^{(a)}\right)\exp\left(\alpha_i^{(b)}\right)\right)=\epsilon(a,b)^{\delta_{a\neq b}}\left(\exp\left(\alpha_i^{(a)}+\alpha_i^{(b)}\right)\right)\quad(i=x,y),\\
		&\left(\exp\left(\alpha_i^{(a)}\right)\exp\left(\alpha_j^{(b)}\right)\right)=\varepsilon(a,b)^{\delta_{a\neq b}}\left(\exp\left(\alpha_x^{(a)}+\alpha_y^{(b)}\right)\right),\quad(i\neq j)\\
		&\left\langle U^{\gamma_0^{(a)}}\exp\left(\widetilde{\gamma}^{(b)}\right)\right\rangle=U^{\delta_{ab}}.
	\end{split}
\end{equation}
Here, $\epsilon(a,b)$ and $\varepsilon(a,b)$ can be any non-zero numbers satisfying $\frac{\epsilon(a,b)}{\epsilon(b,a)}=(-1)^{\chi_{ab}+1}$ and $\frac{\varepsilon(a,b)}{\varepsilon(b,a)}=-1$ (for $a\neq b$). We can then obtain the OPEs for $X^{(a)},Y^{(a)},\Gamma^{(a)}_{\pm}$ from
\begin{equation}
	\begin{split}
		&\left\langle\exp\left(x^{(a)}_+(U)\right)\exp\left(x^{(b)}_-(V)\right)\right\rangle=\prod_i\left(1-\mathfrak{q}H_{ba,i}\frac{V}{U}\right),\\
		&\left\langle\exp\left(y^{(b)}_+(U)\right)\exp\left(y^{(a)}_-(V)\right)\right\rangle=\prod_i\left(1-\mathfrak{q}H_{ab,i}\frac{V}{U}\right)^{-1},\\
		&\left\langle\exp\left(\gamma^{(a)}_+(U)\right)\exp\left(\gamma^{(b)}_-(V)\right)\right\rangle=\left(1-\frac{V}{U}\right)^{-\delta_{ab}}.
	\end{split}
\end{equation}
With these currents, we have
\begin{equation}
	\begin{split}
		&\psi^{(a)}_+(U)=\frac{1}{\mathfrak{q}^{-1}-\mathfrak{q}}\left(\exp\left(x^{(a)}_-\left(\mathfrak{q}^{-1/2}U\right)-x^{(a)}_+\left(\mathfrak{q}^{1/2}U\right)-y^{(a)}_+\left(\mathfrak{q}^{-1/2}U\right)+y^{(a)}_-\left(\mathfrak{q}^{-3/2}U\right)\right)\right),\\
		&\psi^{(a)}_-(U)=\frac{1}{\mathfrak{q}^{-1}-\mathfrak{q}}\left(\exp\left(x^{(a)}_-\left(\mathfrak{q}^{-3/2}U\right)-x^{(a)}_+\left(\mathfrak{q}^{-1/2}U\right)-y^{(a)}_+\left(\mathfrak{q}^{1/2}U\right)+y^{(a)}_-\left(\mathfrak{q}^{-1/2}U\right)\right)\right),\\
		&e^{(a)}(U)=\left(\exp\left(X^{(a)}(U)\right)\partial\exp\left(\Gamma^{(a)}_-(U)\right)\right),\\
		&f^{(a)}(U)=\left(\exp\left(Y^{(a)}(U)\right)\exp\left(\Gamma^{(a)}_+(U)\right)\right).
	\end{split}\label{freefieldrealization}
\end{equation}
This follows from a straightforward check with the use of the property $\delta(Z/W)f(Z)=\delta(Z/W)f(W)$ of the formal delta function for any Laurent series $f(Z)$.

From the above discussions, we can also obtain the free field realization for the elliptic algebras using the dressed operators\footnote{Notice that the dressed operators mentioned in Appendix \ref{serre} are for non-chiral quivers. However, the construction is similar for the chiral cases.}. In other words, $\Psi^{(a)}_{\pm}(U),E^{(a)}(U),F^{(a)}(U)$ have the same expressions as the right hand sides in \eqref{freefieldrealization}.

\section{Discussions}\label{discussions}
Let us mention some properties of the toric quivers that are not discussed above. They should be closely related to the truncations of the quiver BPS algebras, which could lead to important physical consequences.

\paragraph{Specular duality} There is another duality for toric quiver gauge theories known as the specular duality as proposed in \cite{Hanany:2012hi,Hanany:2012vc}. Many concepts and quantities enjoy nice properties under such duality \cite{Cremonesi:2013aba,Bao:2021fjd,Bao:2021gxf}. In general, specular duality does not preserve the mesonic moduli space (except self-dual cases) although the Hilbert series would agree up to some fugacity maps. Instead, it is a duality that preserves the master space \cite{Forcella:2008bb,Forcella:2008eh}. Therefore, we do not expect the quiver BPS algebras to be isomorphic under specular duality. However, it exchanges the internal and external perfect matchings, which are associated to the internal and external points of the toric diagram respectively, of the dual brane tilings.

As each arrow in the quiver can be written in terms of a product of some perfect matchings, the arrows also have a one-to-one correspondence for specular dual theories. It is then natural to wonder if the charge assignments would also follow this correspondence of the arrows. However, we have checked several examples and this is not the case, even for self-dual ones\footnote{For a self-dual quiver, an arrow would often be mapped to a different arrow in the quiver.}. Nevertheless, as argued in \cite{Li:2020rij}, the perfect matchings can be used to determine certain truncations of the quiver Yangian. This is because such truncations come from adding D4-branes to the divisors of the toric CY threefold, which correspond to the lattice points of the toric diagram. In \cite{Li:2020rij}, such truncations were only identified for external (or more precisely, corner) perfect matchings. It could be possible that the truncations from D4-branes associated to internal points can be studied from the specular dual case, where the internal perfect matchings are mapped to the external ones\footnote{Of course, there can also be external lattice points that are not at the corners. Moreover, for non-reflexive polygons, specular duality can relate brane tilings on Riemann surfaces with higher genus \cite{Cremonesi:2013aba}.}.

\paragraph{Deformed VOAs} The truncations of quiver BPS algebras are of particular interest since they are expected to be related to certain vertex operator algebras (VOAs), and hence implement the AGT (aka BPS/CFT) correspondence \cite{Alday:2009aq,Wyllard:2009hg}. Indeed, the truncations of the rational algebras give rise to the (universal enveloping algebras of) rectangular $\mathcal{W}$-algebras \cite{Bao:2022jhy,ueda2022affine,kodera2022coproduct}. We expect that the truncations of the toroidal and even elliptic quiver BPS algebras would lead to deformations of the rational VOAs. In particular, the toroidal algebra for $\mathbb{C}^3$ is shown to be a $\mathtt{q}$-deformation of the $\mathcal{W}_{1+\infty}$-algebra in \cite{Harada:2021xnm}. We conjecture that there exist certain $\mathtt{q}$-deformations of the $\mathcal{W}_{M|N\times\infty}$-algebras such that for toroidal BPS algebras $\mathtt{T}$ associated to the generalized conifolds, we have the following commutative diagram which would give the 5d AGT correspondence:
\begin{equation}
	\includegraphics{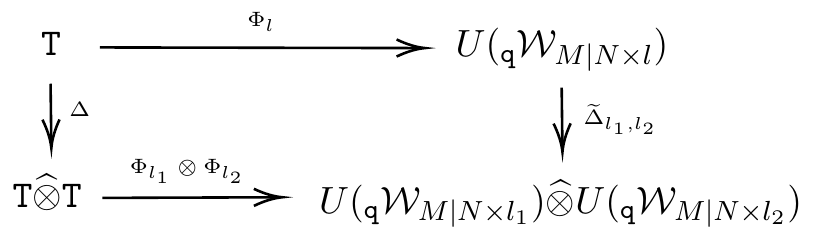},
\end{equation}
where $\Phi_{l}$ are some surjections and the hats denote the completions of the algebras. On the BPS algebra side, this would require a detailed study on the so-called horizontal representations of the algebras with non-trivial central element $C$ so that we can get vertex operators from the generators. On the VOA side, we need to find some suitable deformations of the $\mathcal{W}_{M|N\times\infty}$-algebras studied in \cite{Rapcak:2019wzw,Eberhardt:2019xmf}. It would also be helpful to know more about their free field realizations.

\section*{Acknowledgement}
I am grateful to Christopher Beem, Ian Cheung, Yang-Hui He, Vishnu Jejjala, Jian-Rong Li and in particular, Deshuo Liu, Sasha Ochirov and Rak-Kyeong Seong for helpful discussions. The research is supported by a CSC scholarship.

\appendix

\section{Toric Quiver Gauge Theories}\label{gauge}
In this appendix, we give a quick recap on some properties of 4d $\mathcal{N}=1$ quiver gauge theories from toric geometry. More details can be found in the references mentioned below. See also \cite{Yamazaki:2008bt,He:2016fnb,Bao:2020sqg} for reviews.

\subsection{Toric Duality}\label{toric}
Given a quiver gauge theory with its associated brane tiling, one can study many of its salient features. Let us first consider quivers that are related by Seiberg duality \cite{Seiberg:1994pq} in the toric phase, which is also known as the toric duality \cite{Feng:2000mi,Feng:2001xr,Beasley:2001zp,Feng:2001bn,Feng:2002zw}.

\begin{figure}[h]
	\centering
	\includegraphics[width=10cm]{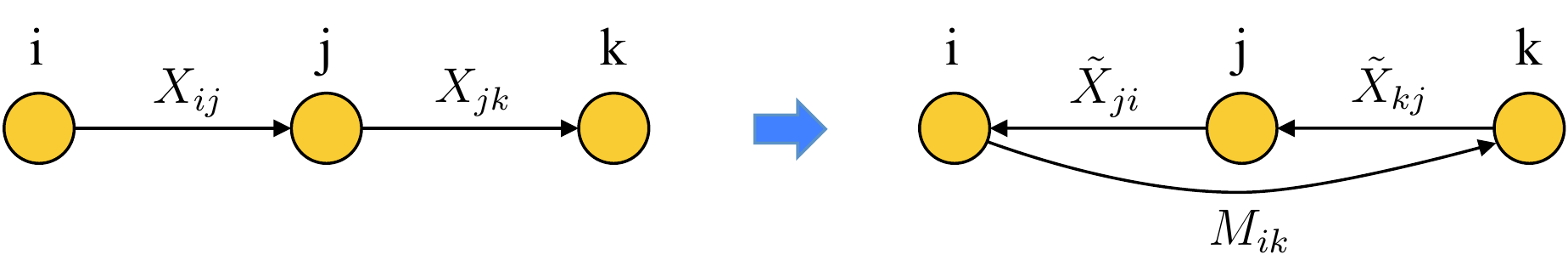}
	\caption{A sketch of how quivers transform under Seiberg duality. Figure taken from \cite{Bao:2020nbi}.}\label{seibergduality}
\end{figure}

In short, picking a node $j$ in the quiver to dualize, we replace all the arrows connected to $j$ with their conjugate (flavour) by reversing their orientations. Then we add a meson, that is, an arrow from $i$ to $k$,  to the new quiver for each 2-path $i\rightarrow j\rightarrow k$ in the original quiver. This process is depicted in Figure \ref{seibergduality}.
In cluster algebra, this is exactly the mutation for quivers (without adjoint loops and 2-cycles) \cite{fomin2002cluster}. The superpotential and the ranks of the gauge groups would be transformed accordingly. In particular, factors $X_{ij}X_{jk}$ should be replaced with $M_{ik}$ in the superpotential, and terms $M_{ik}\widetilde{X}_{kj}\widetilde{X}_{ji}$ need to be added. Moreover, fields that acquire masses, i.e., quadratic terms in the superpotential, should be integrated out using their $F$-term relations \cite{Feng:2001bn}. As we shall consider toric quivers, only nodes with two arrows and two arrows out would be dualized. In terms of brane tilings, this can be performed by the urban renewal \cite{propp2003generalized,Franco:2005rj}.

\subsection{The Higgs-Kibble Mechanism}\label{higgsing}
\begin{figure}[h]
	\centering
	\includegraphics{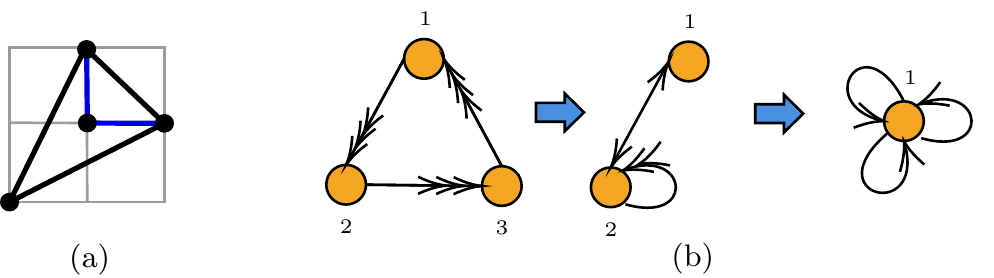}
	\caption{The example of higgsing dP$_0$ (aka $K_{\mathbb{P}^2}$, $\mathbb{C}^3/\mathbb{Z}_3$ $(1,1,1)$) to $\mathbb{C}^3$.}\label{higgsingex}
\end{figure}

As studied in \cite{Feng:2002fv}, higgsing of a theory corresponds to blowing down a compact 2-cycle to a point in the toric geometry while unhiggsing blows up a point to a compact 2-cycle. The process of higgsing can also be nicely encoded in the toric diagrams and in the quivers. An example is depicted in Figure \ref{higgsingex}.

By turning on a VEV of a bifundamental in the quiver at each step, some fields would acquire masses which should be integrated out. Using their equations of motion, we can obtain the superpotential after higgsing similar to the process of performing toric duality. In the above example, the superpotential changes as
\begin{equation}
	\begin{split}
		W=&I_{12}^1I_{23}^3I_{31}^2+I_{12}^1I_{23}^1I_{31}^3+I_{23}^2I_{31}^1I_{12}^3-I_{12}^2I_{23}^3I_{31}^1-I_{12}^1I_{23}^2I_{31}^3-I_{23}^2I_{31}^2I_{12}^3\\
		\overset{\left\langle X_{23}^1\right\rangle=1}{\xrightarrow{\hspace*{1.2cm}}}&I_{22}^1I_{21}^3I_{12}^3I_{22}^2-I_{22}^2I_{21}^3I_{12}^3I_{22}^1\\
		\overset{\left\langle X_{12}^3\right\rangle=1}{\xrightarrow{\hspace*{1.2cm}}}&I_{11}^1\left[I_{11}^3,I_{11}^2\right],
	\end{split}
\end{equation}
where we have omitted the traces.

In terms of brane tilings, turning on a VEV of a bifundamental removes an edge, and integrating out massive fields corresponds to removing (and combining) certain nodes in the dimer. An illustration can be found in \cite[Figure 50]{Hanany:2012hi}.

\subsection{Generalized Conifolds}\label{gencon}
Recall that the toric diagram of any generalized conifold $xy=z^Mw^N$ is a trapezium on the lattice of height one with two horizontal lines of lengths $M$ and $N$. The quiver (in any toric phase) can essentially be viewed as the ``tripled'' quiver\footnote{Here, by ``tripled'', we mean that we first add an opposite arrow for each existing arrow in the Dynkin quiver. Then we only add adjoint loops to the bosonic nodes.} of a Dynkin diagram associated to the underlying untwisted affine Lie superalgebra $\widehat{\mathfrak{sl}}_{M|N}$.

\begin{figure}[h]
	\centering
	\includegraphics{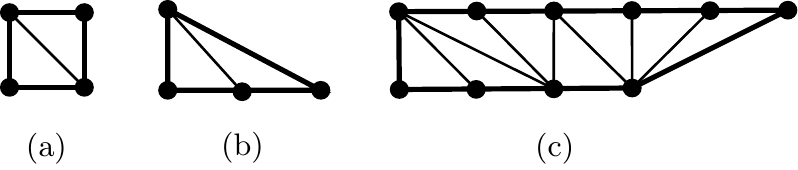}
	\caption{Figure taken from \cite{Bao:2022jhy}. We have (a) $\varsigma=\{-1,+1\}$, (b) $\varsigma=\{-1,-1\}$ and (c) $\varsigma=\{-1,-1,+1,+1,-1,+1,+1,+1\}$.}\label{sigmaex}
\end{figure}

The quivers in different toric phases with different numbers of bosonic and fermionic nodes are encoded by the triangulations of the corresponding toric diagram \cite{nagao2008derived,Nagao:2009rq}. Each simplex in a given triangulation corresponds to a sign $\varsigma_a=\pm1$. Together, they form a parity sequence $\varsigma=\{\varsigma_a|a\in\mathbb{Z}/(M+N)\mathbb{Z}\}$. Overall, the numbers of plus and minus ones are given by $M$ and $N$. If two simplices are aligned side by side, then $\varsigma_a=\varsigma_{a+1}$. If they are aligned in the alternative way, then $\varsigma_a$ and $\varsigma_{a+1}$ have opposite signs. Some illustrations can be found in Figure \ref{sigmaex}.

Besides a pair of opposite arrows connecting each pair of nodes $a$ and $a+1$, the quiver has a self-loop on each bosonic node. If $\varsigma_a=\varsigma_{a+1}$, then the node $a$ is bosonic/even. Otherwise, it is fermionic/odd. The superpotential is composed of terms
\begin{equation}
	\begin{cases}
		\varsigma_a\text{tr}(I_{a,a}I_{a,a-1}I_{a-1,a}-I_{a,a}I_{a,a+1}I_{a+1,a}),&\varsigma_a=\varsigma_{a+1}\\
		\varsigma_a\text{tr}(I_{a,a+1}I_{a+1,a}I_{a,a-1}I_{a-1,a}),&\varsigma_a=-\varsigma_{a+1}
	\end{cases}.
\end{equation}

Following the above rule of toric duality, it is straightforward to see that we can only dualize fermionic nodes in the toric phase. This would just change the parity of the two nodes connected to the dualized node by adding or removing the adjoint loops. Correspondingly, the Dynkin diagrams of the underlying affine Lie superalgebra are related by odd reflections.

A generalized conifold with a larger polygon can be higgsed to one with a smaller polygon. This can be decomposed into a sequence of higgsings. For each single higgsing, the leftmost or rightmost simplex is removed. In the quiver, we merge two adjacent nodes. The two nodes can be either bosonic or fermionic. Suppose that the nodes $a$ and $a+1$ are merged, then $|a'|=|a|+|a+1|$, where $a'$ denotes the corresponding node after higgsing. Let us list how the Cartan matrices would change for the three possible cases:
\begin{align}
	&\includegraphics{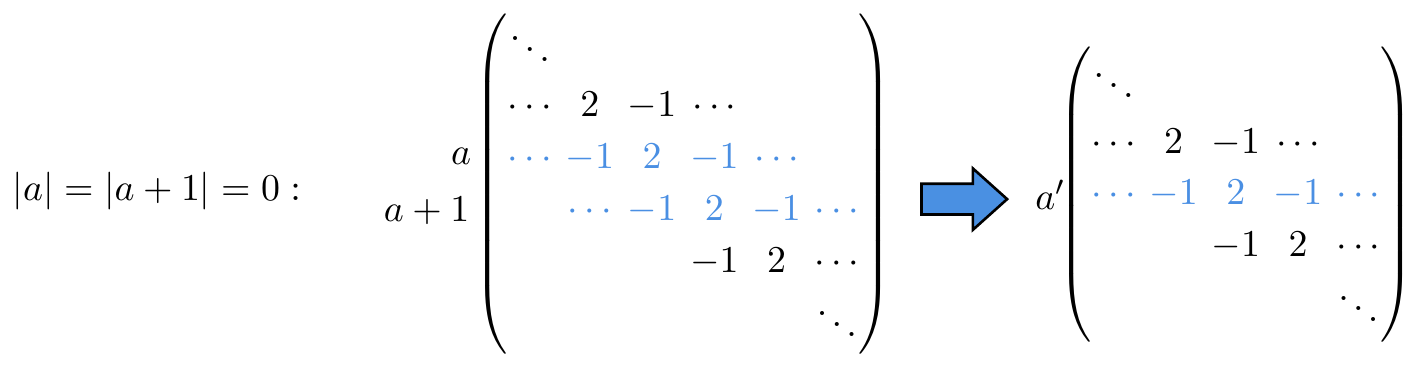},\\
	&\includegraphics{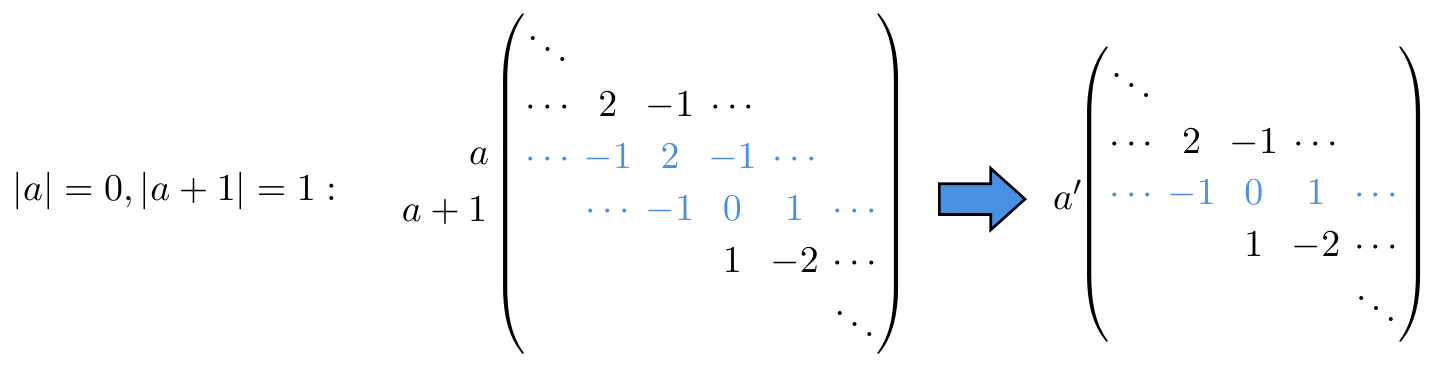},\\
	&\includegraphics{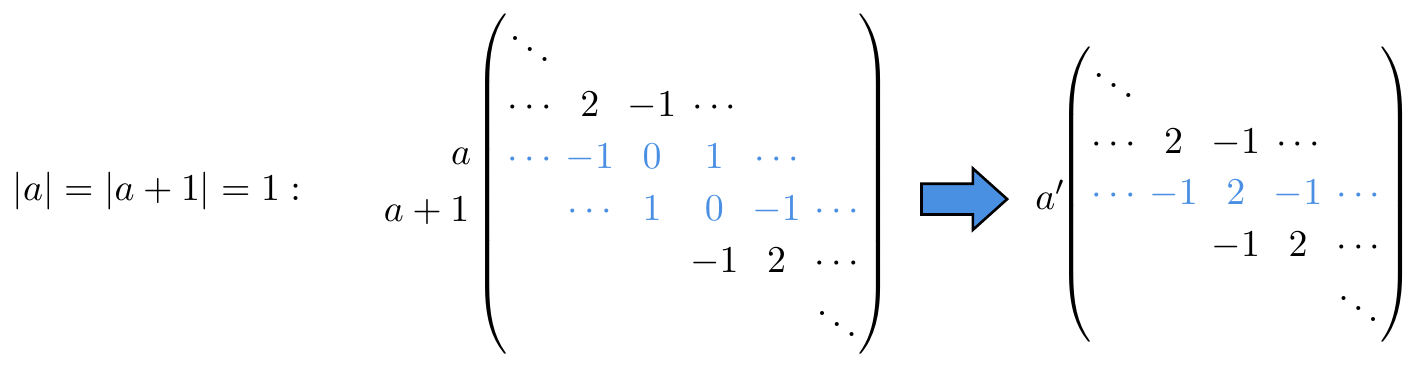}.
\end{align}

\section{Serre Relations}\label{serre}
Besides the relations listed in \S\ref{algebras}, the quiver BPS algebras also have Serre relations. Here, we will only discuss the cases for non-chiral quivers with $M+N\geq3$, $MN\neq2$. Although the Serre relations for general chiral quivers are still not known, examples can be found in \cite{Galakhov:2021vbo}. It is observed that the Serre relations (for either chiral or non-chiral quivers) are closely related to the superpotential of the theory \cite{Li:2020rij}.

For the rational algebras, we have
\begin{align}
	&\underset{u_1,u_2}{\text{Sym}}\left[ e^{(a)}(u_1),\left[ e^{(a)}(u_2),e^{(a\pm1)}(v)\right\}\right\}=0\qquad(|a|=0),\\
	&\underset{u_1,u_2}{\text{Sym}}\left[ e^{(a)}(u_1),\left[ e^{(a+1)}(v_1)\left[ e^{(a)}(u_2),e^{(a-1)}(v_2)\right\}\right\}\right\}=0\qquad(|a|=1),\\
	&\underset{u_1,u_2}{\text{Sym}}\left[ f^{(a)}(u_1),\left[ f^{(a)}(u_2),f^{(a\pm1)}(v)\right\}\right\}=0\qquad(|a|=0),\\
	&\underset{u_1,u_2}{\text{Sym}}\left[ f^{(a)}(u_1),\left[ f^{(a+1)}(v_1)\left[ f^{(a)}(u_2),f^{(a-1)}(v_2)\right\}\right\}\right\}=0\qquad(|a|=1).
\end{align}
For the toroidal algebras, the Serre relations are
\begin{align}
	&\underset{u_1,u_2}{\text{Sym}}\left\llbracket e^{(a)}(u_1),\left\llbracket e^{(a)}(u_2),e^{(a\pm1)}(v)\right\rrbracket_{H_1}\right\rrbracket_{H_1}=0\qquad(|a|=0),\\
	&\underset{u_1,u_2}{\text{Sym}}\left\llbracket e^{(a)}(u_1),\left\llbracket e^{(a+1)}(v_1)\left\llbracket e^{(a)}(u_2),e^{(a-1)}(v_2)\right\rrbracket_{H_1}\right\rrbracket_{H_1}\right\rrbracket_{H_1}=0\qquad(|a|=1),\\
	&\underset{u_1,u_2}{\text{Sym}}\left\llbracket f^{(a)}(u_1),\left\llbracket f^{(a)}(u_2),f^{(a\pm1)}(v)\right\rrbracket_{H_1^{-1}}\right\rrbracket_{H_1^{-1}}=0\qquad(|a|=0),\\
	&\underset{u_1,u_2}{\text{Sym}}\left\llbracket f^{(a)}(u_1),\left\llbracket f^{(a+1)}(v_1)\left\llbracket f^{(a)}(u_2),f^{(a-1)}(v_2)\right\rrbracket_{H_1^{-1}}\right\rrbracket_{H_1^{-1}}\right\rrbracket_{H_1^{-1}}=0\qquad(|a|=1).
\end{align}
Here, the $\mathfrak{q}$-graded bracket is given by $\llbracket x,y\rrbracket_{\mathfrak{q}}=xy-(-1)^{|x||y|}\mathfrak{q}^{(x,y)}yx$, where $(x,y)$ is the root pairing stemmed from the underlying affine Lie superalgebra. For instance, the pairing of two simple roots gives the corresponding entry in the Cartan matrix.

As we can see, both of the two types of the algebras have their versions of the brackets. Therefore, we would also like to use an ``elliptic bracket'' to write the Serre relations for the elliptic cases. Let us introduce the operators $\chi_a(u)$ and $\xi_a(u)$ that commute with all $e$, $f$, $\psi_{\pm}$ generators in the elliptic algebras. They have the following correlators:
\begin{align}
	&\text{e}^{\langle\chi_a(u)\chi_b(v)\rangle}=\frac{\left(qH_1^{A_{ab}}H_2^{-M_{ab}}U^{-1}V;q\right)_{\infty}}{\left(qH_1^{-A_{ab}}H_2^{-M_{ab}}U^{-1}V;q\right)_{\infty}},\\
	&\text{e}^{\langle\xi_a(u)\xi_b(v)\rangle}=\frac{\left(qH_1^{-A_{ab}}H_2^{-M_{ab}}U^{-1}V;q\right)_{\infty}}{\left(qH_1^{A_{ab}}H_2^{-M_{ab}}U^{-1}V;q\right)_{\infty}},\\
	&\text{e}^{\langle\chi_a(u)\xi_b(v)\rangle}=1.
\end{align}
Then using the correlators of the ``dressed'' operators
\begin{equation}
	E^{(a)}(u)=\text{e}^{\chi_a(u)}e^{(a)}(u),\quad F^{(a)}(u)=\text{e}^{\xi_a(u)}f^{(a)}(u),\quad\Psi^{(a)}_{\pm}(u)=\text{e}^{\chi_a(u\pm c/2)}\text{e}^{\xi_a(u\mp c/2)}\psi^{(a)}_{\pm}(u),\label{dressedoperators}
\end{equation}
the relations of the elliptic algebras can be written in the same forms as those of the toroidal algebras. For instance, the $ee$ relations of the elliptic algebras now become
\begin{equation}
	\left(H_2^{M_{ab}}U-H_1^{A_{ab}}V\right)\left\langle E^{(a)}(u)E^{(b)}(v)\right\rangle=(-1)^{|a||b|}\left(H_1^{A_{ab}}H_2^{M_{ab}}U-V\right)\left\langle E^{(b)}(v)E^{(a)}(u)\right\rangle.
\end{equation}
Therefore, the Serre relations of the elliptic algebras can simply be obtained by taking the ones of the toroidal algebras. Then we replace the toroidal generators with the dressed elliptic generators and take the correlators of the whole expressions. For brevity, we shall write them using the ``elliptic brackets'' as
\begin{align}
	&\underset{u_1,u_2}{\text{Sym}}\left[ e^{(a)}(u_1),\left[ e^{(a)}(u_2),e^{(a\pm1)}(v)\right\}_{\chi}\right\}_{\chi}=0\qquad(|a|=0),\\
	&\underset{u_1,u_2}{\text{Sym}}\left[ e^{(a)}(u_1),\left[ e^{(a+1)}(v_1)\left[ e^{(a)}(u_2),e^{(a-1)}(v_2)\right\}_{\chi}\right\}_{\chi}\right\}_{\chi}=0\qquad(|a|=1),\\
	&\underset{u_1,u_2}{\text{Sym}}\left[ f^{(a)}(u_1),\left[ f^{(a)}(u_2),f^{(a\pm1)}(v)\right\}_{\xi}\right\}_{\xi}=0\qquad(|a|=0),\\
	&\underset{u_1,u_2}{\text{Sym}}\left[ f^{(a)}(u_1),\left[ f^{(a+1)}(v_1)\left[ f^{(a)}(u_2),f^{(a-1)}(v_2)\right\}_{\xi}\right\}_{\xi}\right\}_{\xi}=0\qquad(|a|=1).
\end{align}

\section{Conventions of Heisenberg Modes}\label{kmodes}
In the main context, we introduced the modes $k_r$ (and $l_r$) for the $\psi_{\pm}$ currents. Here, we mention some alternative convention to define these Heisenberg modes. It could be possible that this would be more convenient when considering certain aspects of the algebras such as their representations and the AGT correspondence.

Let us consider the toroidal algebras for non-chiral quivers as an example. The other cases can be redefined in a similar manner. First, we rescale the $e$, $f$ modes as
\begin{equation}
	e^{(a)}_n=\left(\mathtt{q}-\mathtt{q}^{-1}\right)^{1/2}\mathtt{e}^{(a)}_n,\quad f^{(a)}_n=\left(\mathtt{q}-\mathtt{q}^{-1}\right)^{1/2}\mathtt{f}^{(a)}_n,
\end{equation}
where we have suggestively written $\mathtt{q}=\exp(\beta h_1)=H_1$. Notice that this does not change the $ee$ and $ff$ relations. Then the $e_0f_0$ relations (as well as the $e_nf_{-n}$ relations) would become
\begin{equation}
	\left[\mathtt{e}^{(a)}_0,\mathtt{f}^{(a)}_0\right\}=\delta_{ab}\frac{\mathtt{q}^{k^{(a)}_0}-\mathtt{q}^{-k^{(a)}_0}}{\mathtt{q}-\mathtt{q}^{-1}}=\delta_{ab}\left[k^{(a)}_0\right]_{\mathtt{q}}.
\end{equation}
Here, $[x]_{\mathtt{q}}=\frac{\mathtt{q}^{x}-\mathtt{q}^{-x}}{\mathtt{q}-\mathtt{q}^{-1}}$ is the standard $\mathtt{q}$-number. On the other hand, the $k_0\mathtt{e}_n$ (resp. $k_0\mathtt{f}_n$) relations remain the same as the ones for $k_0e_n$ (resp. $k_0f_n$). As we can see, the relations among the zero modes resemble the ones appeared in quantum groups.

Likewise, we can write
\begin{equation}
	\psi^{(a)}_{\pm}(U)=\psi^{(a)}_{\pm,0}\exp\left(\left(\mathtt{q}-\mathtt{q}^{-1}\right)\sum_{n=0}^{\infty}\mathtt{k}^{(a)}_{\pm n}U^{\mp n}\right)
\end{equation}
such that $k^{(a)}_r=\left(\mathtt{q}-\mathtt{q}^{-1}\right)\mathtt{k}^{(a)}_r$. Therefore,
\begin{equation}
	\psi^{(a)}_{\pm,n}=\psi^{(a)}_{\pm,0}\sum_{m=1}^n\frac{\left(\mathtt{q}-\mathtt{q}^{-1}\right)^m}{m!}\sum_{\substack{r_1,\dots,r_m>0\\r_1+\dots+r_m=n}}\mathtt{k}^{(a)}_{\pm r_1}\mathtt{k}^{(a)}_{\pm,r_2}\dots\mathtt{k}^{(a)}_{\pm,r_m}.
\end{equation}
The commutation relations involving $\mathtt{k}^{(a)}_r$ can be obtained with the substitutions
\begin{equation}
	H_1^{rA_{ab}}-H_1^{-rA_{ab}}\rightarrow[rA_{ab}]_{\mathtt{q}},\quad C^{-r}-C^r\rightarrow\frac{C^{-r}-C^r}{\mathtt{q}-\mathtt{q}^{-1}}=-[rc/h_1]_{\mathtt{q}}
\end{equation}
in the relations for $k^{(a)}_r$.

Sometimes, it is also conventional to define the Heisenberg modes with signs inside the exponentials. In other words, we have $\exp\left(\pm\sum\limits_nk_{\pm n}U^{\mp n}\right)$ in the expressions for $\psi_{\pm}$. This is simply a redefinition of $k_{-n}\rightarrow-k_{-n}$.

\addcontentsline{toc}{section}{References}
\bibliographystyle{utphys}
\bibliography{references}

\end{document}